\documentclass[a4paper, notoc]{JHEP3}
\usepackage[dvips]{graphicx}
\usepackage{amsfonts}
\usepackage{amssymb}
\usepackage{epsfig}
\usepackage{cite}

\newcommand{\be}{\begin{equation}}
\newcommand{\ee}{\end{equation}}
\newcommand{\bea}{\begin{eqnarray}}
\newcommand{\eea}{\end{eqnarray}}
\newcommand{\mbb}{\mathbb}
\newcommand{\ti}{\times}
\newcommand{\half}{\frac{1}{2}}
\newcommand{\mc}{\mathcal}

\newcommand{\beqa}{\begin{eqnarray}}
\newcommand{\eeqa}{\end{eqnarray}}

\newcommand{\ph}{\phantom}

 \title{Gauge Threshold Corrections for Local String Models}
\author{Joseph P. Conlon
 \\ Rudolf Peierls Center for Theoretical Physics, 1 Keble Road \\
 Oxford OX1 3NP, UK }

\abstract{We study gauge threshold corrections for local brane models embedded in 
a large compact space. A large bulk volume gives important contributions to 
the Konishi and super-Weyl anomalies and the effective field theory analysis implies the 
unification scale should be enhanced in a model-independent way from $M_s$ to $R M_s$. 
For local D3/D3 models this result is supported by the explicit string computations. In this 
case the scale $R M_s$ comes from the necessity of global cancellation of RR tadpoles sourced by the local model. 
We also study D3/D7 models and discuss discrepancies with the effective field theory analysis.
We comment on phenomenological implications for gauge coupling unification and for 
the GUT scale.}

\preprint{OUTP-09/03P}

\begin{document}

\section{Introduction}

One of the most tantalising results in particle physics is the appearance of
gauge coupling unification at an energy scale $M_{GUT} \sim 10^{16} \hbox{GeV}$
in the presence of low energy supersymmetry. 
It is possible this unification is simply an accident; however, it may instead be the first hint of 
a deeper structure underlying the Standard Model. Unfortunately, it is difficult to
probe such high energies directly and we do not know whether $M_{GUT}$ does indeed mark a new physical scale.

String theory is the most promising candidate for an ultraviolet theory containing a unified treatment of gauge and
gravitational interactions. Model building and unification in string theory was traditionally studied in the context of the
heterotic string, which naturally gives rise to grand unified theories at the compactification scale.
However more recently much model building has shifted to intersecting brane models in type II theories (see \cite{0610327} for a 
review). Intersecting brane models naturally give chiral fermions and $SU(n)$ nonabelian gauge groups.
However, in general these models do not give gauge coupling unification except in certain special limits.

One of these limits is the case of branes at singularities, where the dilaton provides the universal gauge coupling.
This limit is a particular case of the class of local models, first introduced in \cite{aiqu}.
String models can be classified as either global or local. In global models, for example
the weakly coupled heterotic string, the string scale is tied to the Planck scale: if $M_s \ll M_P$ then $\alpha_i \ll 1$ in a way
incompatible with phenomenology. In local models the string scale and Planck scale can 
be decoupled: the Standard Model gauge couplings have no direct relation to the ratio $M_P/M_s$.

Local models are attractive in that they fit naturally into, and are required by, controlled 
scenarios of moduli stabilisation with low scale supersymmetry breaking 
\cite{hepth0502058, hepth0505076}. However a tension arises between gauge coupling unification
and supersymmetry breaking. In flux-stabilised models, the most attractive value of the string scale 
with regards to supersymmetry breaking is $M_s \sim 10^{11} \hbox{GeV}$ (for early discussions of an intermediate 
string scale see \cite{9809582, 9810535}).
This solves the hierarchy problem through TeV scale supersymmetry at $M_{susy} \sim M_s^2/M_P$. This scale is also attractive
with regard to axions and neutrino masses. However, this leads us to expect gauge couplings to unify at a scale $M_s$ rather than 
$M_{GUT} \sim 10^{16} \hbox{GeV}$.

The purpose of this paper is study this tension and in particular the effects of threshold corrections 
on the unification scale for local models.\footnote{A related study has recently been carried out in
\cite{08082223}, although focusing on the contribution of local modes rather than the dependence on the bulk volume.}
 A local model embedded in a compact space naturally has a large parameter, the 
ratio $(M_P/M_s)^2 = \mc{V}$.\footnote{Volumes will be treated as dimensionless
and measured in units of $(2 \pi \sqrt{\alpha'})$: so $\mc{V} = \hbox{Vol}/(2 \pi \sqrt{\alpha'})^6$}
This large parameter enters both the K\"ahler potential and the matter kinetic terms. However these are both
known to modify the physical gauge coupling through the Konishi and super-Weyl anomalies. Although formally one-loop effects,
both anomalies will be parametrically enhanced at large volume and so may lead to significant effects on the gauge couplings.

This paper is organised as follows. Section \ref{secFT} studies the effects of threshold corrections from an effective
field theory perspective, and shows that this leads us to expect unification at a super-stringy scale
$M_X = R M_s$. The remainder of this paper studies threshold corrections from a stringy perspective. Section 3
reviews the formalism for computing threshold corrections, which is applied in section 4 to local D3/D3 models and in section 5
to D3/D7 models. Section 6 contains the conclusions, while an appendix contains various useful $\vartheta$-function identities.

\section{Field Theory Results}
\label{secFT}

The defining characteristic of local models is that there exists a limit in which gravity decouples - the ratio
$M_P/M_s$ can be taken to infinity without affecting Standard Model gauge and Yukawa couplings.
In phenomenological applications, the bulk volume may be very large: for example in the large volume scenario
of \cite{hepth0502058, hepth0505076} the volume is $\mc{V} \sim 10^{15} l_s^6$. For the GUT-like models
of \cite{08023391, 08060102, 08112936} the proposed volume is $\mc{V} \sim 10^4$.
As this large number enters into both the overall K\"ahler potential and the matter kinetic terms, 
it is important to study the anomaly-induced corrections to gauge couplings.\footnote{For related studies of gauge coupling
unification in the phenomenological literature on extra dimensions, see \cite{9908146, 9908530}, and for early studies of
threshold corrections in the presence of large extra dimensions see \cite{antold}.}

In locally supersymmetric effective field theory with field-dependent couplings
the physical gauge couplings are given by the Kaplunovsky-Louis formula \cite{9303040, 9402005}:
\bea
\label{KL}
g_a^{-2}(\Phi, \bar{\Phi}, \mu) & = & \hbox{Re}(f_a(\Phi)) + \frac{\left( \sum_r n_r T_a(r) - 3T_a(G)\right)}{8 \pi^2}
\ln \left( \frac{M_P}{\mu}\right) + \frac{T(G)}{8 \pi^2} \ln g_a^{-2} (\Phi, \bar{\Phi}, \mu)
\nonumber \\ 
& & + \frac{(\sum_r n_r T_a(r) - T(G))}{16 \pi^2} \hat{K}(\Phi, \bar{\Phi}) 
 - \sum_r \frac{T_a(r)}{8 \pi^2} \ln \det Z^r(\Phi, \bar{\Phi}, \mu). 
\eea
$\Phi$ are the moduli, $\hat{K}$ is the moduli K\"ahler potential, and $Z^r$ the K\"ahler metric for matter
in representation $r$. We will focus on the volume dependence of (\ref{KL}) and in particular 
on terms enhanced by $(\ln \mc{V})$ in the large volume limit,
and therefore drop the NSVZ term $\frac{T(G)}{8 \pi^2} \ln g_a^{-2}$ in all 
subsequent formulae.

The KL formula (\ref{KL}) relates the physical and holomorphic gauge couplings in locally supersymmetric theories,
generalising the NSVZ formula for globally supersymmetric theories. The left hand side contains the physical couplings.
On the right hand side, the first term is the holomorphic gauge coupling and the second represents the standard field theory running.
The third term is the NSVZ relationship between physical and holomorphic couplings, whereas the fourth is the specifically
supergravity contribution from the super-Weyl anomaly. This originates from the need to transform from Weyl to Einstein frame when
relating couplings with manifest holomorphy properties to couplings with a direct physical interpretation. The last term is the Konishi
anomaly associated with rescaling matter fields to canonical normalisation. 

The moduli dependence of (\ref{KL}) originates from $\hat{K}$ and $Z^r$, which in string theory are functions
of the moduli. As the anomalies are already one-loop effects, to compute (\ref{KL}) to 
one-loop level, it is sufficient to know $\hat{K}$ and $Z^r$ at tree level. The overall K\"ahler potential
$\hat{K}$ is given by \cite{0403067} (we neglect any additional dependence on brane/Wilson line moduli)
\be
\label{kpot}
\hat{K} = - 2 \ln (\mc{V}(T_i + \bar{T}_i)) - \ln \left(\int i \Omega \wedge \bar{\Omega} \right) - \ln (S + \bar{S}).
\ee
The last two terms of (\ref{kpot}) depend on dilaton ($S$) and complex structure ($U$) moduli and are not relevant to our
purposes.
The form of $\hat{K}$ can be easily understood from the supergravity scalar potential, 
$V = e^{\hat{K}} \left( \vert D_I W \vert^2 - 3\vert W \vert^2 \right)$.
String theory dimensional analysis requires $V \sim M_s^4 = \frac{M_P^4}{\mc{V}^2}$, 
requiring $e^K \sim \frac{1}{\mc{V}^2}$.

The form of $Z^r$ can be deduced using the shift symmetries of the K\"ahler moduli. In a local model 
the physical Yukawas
\be
\hat{Y}_{\alpha \beta \gamma} = \frac{e^{\hat{K}/2} Y_{\alpha \beta \gamma}}{\sqrt{Z_{\alpha} Z_{\beta} Z_{\gamma}}}
\ee
must necessarily be independent of $\mc{V}$. However the superpotential Yukawas $Y_{\alpha \beta \gamma}$ 
must - at least perturbatively - be independent of $T_i$, as $Y_{\alpha \beta \gamma}$ must both
be holomorphic in $T_i$ and respect the shift symmetry $T_i \to T_i + \epsilon$. This requires
\be 
\sqrt{Z_{\alpha} Z_{\beta} Z_{\gamma}} \sim \frac{1}{\mc{V}}.
\ee
If we assume that local fields see the bulk volume in the same way, we obtain 
$Z_{\alpha} \sim \frac{1}{\mc{V}^{2/3}}$. In models where the fields are related by symmetries, this assumption automatically
holds. This is the scaling of the kinetic term for $33$ strings in a bulk space. 

In this case $\hat{K}$ and $Z^r$ are given by
$$ 
\hat{K} = - 2 \ln \mc{V} = - 4 \ln \left( \frac{M_P}{M_s} \right), \qquad Z = \frac{1}{\mc{V}^{2/3}}, \qquad \ln \det Z = - \frac{4 n_r}{3} \ln %%@
\left( \frac{M_P}{M_s} \right). 
$$ 
We therefore obtain (writing $M_X = \left( \frac{M_P}{M_S} \right)^{1/3} M_S$),
\bea
g_a^{-2}(\mu) & = & \hbox{Re}(f_a(\Phi)) + \left( \frac{\sum_r n_r T_a(r) - 3T_a(G)}{16 \pi^2} \right) \ln \left( \frac{M_P^2}{\mu^2} \right) %%@
\nonumber \\
& & - \frac{2}{16 \pi^2} \left( \sum_r n_r T_a(r) - T(G) \right) \ln \left( \frac{M_P^2}{M_s^2} \right)
 - \left( \sum_r \frac{- 4 n_r T_a(r)}{3 \ti 16 \pi^2} \right) \ln \left(  \frac{M_P^2}{M_s^2} \right) \nonumber \\
 & = & \hbox{Re}(f_a(\Phi)) + \ln \left( \frac{M_X^2}{\mu^2} \right) \left( \frac{\sum_r n_r T_a(r) -3 T_a(G)}{16 \pi^2} \right) \nonumber \\
& = & \hbox{Re}(f_a(\Phi)) + \beta_a \ln \left( \frac{M_X^2}{\mu^2} \right).
\eea
This expression implies the effect of the 
K\"ahler and Konishi anomalies is to modify the naive unification scale of $M_S$ and raise it to
a scale $M_X$, where
\be
\label{mir}
M_X = \left( \frac{M_P}{M_S} \right)^{1/3} M_S.
\ee
$M_X$ is enhanced compared to the string scale by a factor of the compactification radius.
This effect is substantial; for intermediate string scale models it moves the naive unification scale
from a range $10^{11} \hbox{GeV} \div 10^{12} \hbox{GeV}$ to a range $2.5 \ti 10^{13} \hbox{GeV} \div
1.5 \ti 10^{14} \hbox{GeV}$.  

We assumed here that $\hbox{Re}(f_a(\Phi))$ is gauge group universal, in order that
non-accidental unification may make sense in the first place. This is realised for example by
models of D3 branes on a singularity, where $\hbox{Re}(f_a(\Phi)) = S$.
In many intersecting brane models $\hbox{Re}(f_a(\Phi))$ is far from universal, and in this case
the unification of gauge couplings must necessarily be accidental. Mirage unification may also occur if
the non-universality of $\hbox{Re}(f_a)$ is related to the $\beta$-functions. For a discussion of this possibility, see
\cite{9905349, 9906039}.

The surprising feature of (\ref{mir}) is the simplicity of the calculations that have led to it.
The only assumption made has been that different matter fields see the bulk volume in the same way, which does not
seem a strong assumption for local models where the bulk can in principle be decoupled.
Unification at a scale above the string scale has then followed only from the large volume behaviour
of the K\"ahler potential and simple assumptions about locality. In particular, the above arguments
are independent of the detailed form of the local model, and use only the scaling behaviour 
with the volume. 

A more general point is that the form of (\ref{KL}) implies that threshold corrections 
are non-neglible in any local model. The presence of a large bulk volume, essential for 
the concept of a local model, gives significant $(\ln \mc{V})$ factors 
that must be taken into account in any comparison
with gauge coupling unification.

However, the use of the field theory expressions for anomalous contributions to gauge couplings often has subtle features
such as field redefinitions and chiral/linear multiplet relations.   
In the rest of this paper we therefore set out to study the threshold corrections from a
directly stringy perspective, in order to understand the appearance of $\ln \mc{V}$ terms and 
the apparently model-independent form of (\ref{mir}).
 
\section{Threshold Corrections in String Theory}

The study of threshold corrections in string theory has a long history. The original calculations were carried
out for the heterotic string \cite{9205068, dkl, dkl2, hepth9505046, ant} (see \cite{hepth9602045, 9709062} for reviews). With the advent of brane %%@
models threshold corrections have also
been computed for D-brane models \cite{9605028, 9807082, 9807415, 9906039, 0204094, 0302221, 0612234, 07052150, 08051874}.
This paper will make most use of the presentation given in \cite{9906039}. Let us also state at this point that throughout this paper
all calculations will be carried out in the orbifold limit, where the tree-level gauge couplings are universal.

Threshold corrections in string theory are most straightforwardly computed using the
background field method. Threshold corrections to gauge group $a$ are found by
computing the vacuum energy in a background magnetic field $F_{23} = Q_a B$, 
with $Q_a$ a generator of the gauge group, and extracting the $\mc{O}(B^2)$ contribution. 
In string theory the one-loop vacuum energy involves a sum over
Torus, Klein bottle, Mobius Strip and Annulus diagrams.
\be
\label{vacen}
\Lambda_{1-loop} = \half (T + KB + A(B) + MS(B)).
\ee
As only open strings couple to the magnetic field, only the
annulus and Mobius strip diagrams can depend on $B$ and thus contribute to threshold
corrections.

The vacuum energy has the form
\be
\Lambda = \Lambda_{0} + \half \left( \frac{B}{2 \pi^2} \right)^2 \Lambda_2 + \frac{1}{4!} \left( \frac{B}{2 \pi^2} \right)^4 \Lambda_4 + \ldots
\ee
$\Lambda_0$ vanishes in a supersymmetric compactification. $\Lambda_2$ takes the form
\be
\label{trk}
\Lambda_2 = \int_{0}^{\infty} \frac{dt}{8t} \, f_{thresh}(q \equiv e^{-\pi t}).
\ee
The physical gauge couplings are given by
\be
\frac{4 \pi^2}{g_a^2} \Bigg\vert_{\textrm{1-loop}} = \frac{4 \pi^2}{g_a^2} \Bigg\vert_{\textrm{tree}} + 2 \Lambda_2.
\ee
The threshold corrections are encoded in $f_{thresh}$. In the IR limit $t \to \infty$, $f_{thresh} \to b_a$, where $b_a$
is the field theory beta function coefficient. The $t \to \infty$ limit therefore gives the standard low-energy field theory 
running of the gauge coupling. $t=1$ corresponds to the turn-on of stringy physics, where $e^{-m^2 t \alpha'} \gtrsim 1$.
The stringy physics is encoded in the UV $t \to 0$ limit. In a consistent compact model,
(\ref{trk}) will be finite in the
$t \to 0$ limit. This finiteness is equivalent to the global consistency of the string theory, 
namely the cancellation of all RR tadpoles.

Local models can also have a further simplification, which will hold for the cases considered below. 
At a singularity supersymmetric branes can carry both positive and negative RR charge. 
The cancellation of (local) twisted tadpoles can then be achieved solely using branes and does not require the 
presence of orientifold planes. In this case the M\"obius amplitude is also absent and we can restrict to
considering solely the annulus amplitude.
The annulus amplitude is the partition function for all open string states in the spectrum.
For a $\mbb{Z}_K$ singularity it is given by  
\bea
\mc{A} & = & \int_0^{\infty} \frac{dt}{2t} \, \hbox{STr}\left( \frac{(1 + \theta + \theta^2 + \ldots + \theta^{K-1})}{K}
\frac{1 + (-1)^F}{2} \, q^{(p^\mu p_\mu + m^2)/2} \right) \\
& \equiv & \sum_{i=0}^{K-1} \quad \int_0^{\infty} \frac{dt}{2t} \quad \frac{\mc{A}^{k}}{K}.
\eea
Here $q=e^{-\pi t}$ and $\hbox{STr} = \sum_{bosons} - \sum_{fermions} \equiv \sum_{NS} - \sum_R$. We have also set
$\alpha' = 1/2$.

In this paper we shall perform calculations in local models, without providing an explicit embedding into compact models.
Such an embedding is of course necessary for consistency and for cancellation of all RR tadpoles, but is also model
dependent. The absence of a compact embedding 
means that our computation of threshold corrections will be incomplete. In particular, we will be 
missing states corresponding 
to strings stretching from the local singularity to branes/O-planes in the bulk. However, such states 
have masses $m^2 \sim R^2/((2 \pi)^2 \alpha')$ and will only have non-negligible contributions to the partition function for
$t \lesssim 1/R^2$. As we shall discuss in greater detail below, 
all local results are therefore reliable for $t \gtrsim 1/R^2$ but should be cut off at $t \sim 1/R^2$.

We shall study both D3/D3 and D3/D7 models. We first provide the formalism that is common to all cases, 
before specialising to individual models.
We start by writing the partition functions for the various sectors in the absence of a 
space-time magnetic field.
The purpose of this is primarily to review formulae and to define notation and
conventions. 

\subsection{Unmagnetised Amplitudes}
\label{subsecunmag}

\subsubsection{D3-D3 Amplitudes}

The untwisted D3-D3 annulus amplitude is
\be
\mc{A}_{33}^{0} = \int \frac{dt}{2t} \frac{1}{(2 \pi^2 t)^2} \hbox{Tr}(1)\vert_{CP} \sum_{\alpha, \beta=0,1/2} 
\frac{\eta_{\alpha \beta}}{2} 
\frac{\left( \vartheta \Big[ \begin{array}{c} \alpha \\ \beta \end{array} \Big] \right)^4}{\eta^{12}}.
\ee
The trace is over Chan-Paton states and the sum over $\alpha$ and $\beta$ reflects the 
GSO projection and supertrace, with $\eta_{\alpha \beta} = (-1)^{2(\alpha + \beta - 2 \alpha \beta)}$.
 $\alpha = 0 (1/2)$ corresponds to NS (R) states, and $\beta = 0 (1/2)$ corresponds to
the insertion of $1 ((-1)^F)$ in the trace.

For a fully twisted sector ($\theta_i \neq 0, i=1,2,3$), the twisted D3-D3 partition function is
\be
\mc{A}_{33}^{(k)} = \int \frac{dt}{2t} \frac{1}{(2 \pi^2 t)^2} \hbox{Tr}(\gamma_{\theta^k} \otimes \gamma^{-1}_{\theta^k})\vert_{CP} 
\sum_{\alpha, \beta=0,1/2}  \frac{\eta_{\alpha \beta}}{2}
\frac{\vartheta \Big[ \begin{array}{c} \alpha \\ \beta \end{array} \Big]}{\eta^3}
\prod_{i=1}^3 \left( - 2 \sin \pi \theta_i \right)
\frac{\vartheta \Big[ \begin{array}{c} \alpha \\ \beta + \theta_i \end{array} \Big]}
{\vartheta \Big[ \begin{array}{c} 1/2 \\ 1/2 + \theta_i \end{array} \Big] }.
\ee
For a partially twisted sector ($\theta_i \neq 0, i=1,2$), the twisted D3-D3 partition function is
\be
\mc{A}_{33}^{(k)} = \int \frac{dt}{2t} \frac{1}{(2 \pi^2 t)^2} \hbox{Tr}(\gamma_{\theta^k} \otimes \gamma^{-1}_{\theta^k})\vert_{CP} 
\sum_{\alpha, \beta=0,1/2}  \frac{(-1)^{2 \alpha} \eta_{\alpha \beta}}{2}
\Bigg(\frac{\vartheta \Big[ \begin{array}{c} \alpha \\ \beta \end{array} \Big]}{\eta^3}\Bigg)^2
\prod_{i=1}^2 \left( - 2 \sin \pi \theta_i \right)
\frac{\vartheta \Big[ \begin{array}{c} \alpha \\ \beta + \theta_i \end{array} \Big]}
{\vartheta \Big[ \begin{array}{c} 1/2 \\ 1/2 + \theta_i \end{array} \Big] }.
\ee
The above amplitudes automatically vanish due to supersymmetry. However, in a consistent theory the NSNS
and RR tadpoles must vanish separately once the amplitude is rewritten in closed string tree channel.
The cancellation of closed string RR tadpoles constrains the matter content of the theory and is equivalent to
anomaly cancellation.

To transform to closed string channel, we write $t=1/l$ and use the modular properties
(\ref{EtaTransform}) and (\ref{ThetaTransform}), before extracting the $l \to \infty$ divergence. 
The fully twisted amplitude becomes
\be
\hbox{Tr}(\gamma_{\theta}) \hbox{Tr}(\gamma_{\theta}^{-1}) \int \frac{dl}{(2 \pi^2)^2} \sum \frac{\eta_{\alpha \beta}}{2}
\frac{\vartheta \Big[ \begin{array}{c} \beta \\ \alpha \end{array} \Big]}{\eta^3}
\prod_{i=1}^3 \left( - 2 \sin \pi \theta_i \right)
\frac{e^{2 \pi i \alpha(\beta + \theta_i)} \vartheta \Big[ \begin{array}{c} -\beta - \theta_i \\ \alpha \end{array} \Big]}
{e^{\pi i(1/2 + \theta_i)} \vartheta \Big[ \begin{array}{c} -1/2 - \theta_i \\ 1/2 \end{array} \Big] }(l).
\ee
This generates a twisted tadpole in the RR sector, with the divergence given by
\be
(1_{NSNS} - 1_{RR}) \hbox{Tr}(\gamma_{\theta}) \hbox{Tr}(\gamma_{\theta}^{*}) \int \frac{2 dl}{(2 \pi^2)^2}
\prod_{i=1}^3 \left( - 2 \sin \pi \theta_i \right).
\ee
In the absence of D7 branes this requires $\hbox{Tr}(\gamma_{\theta})=0$.

\subsubsection{D3-D7 amplitudes}

The untwisted D3-D7 amplitudes are
\bea
\mc{A}_{37}^{(0)} & = & \int \frac{dt}{2t} \frac{1}{(2 \pi^2 t)^2} (\hbox{Tr}({ \bf 1}) + \hbox{Tr}({\bf 1})) 
\sum_{\alpha, \beta=0,1/2}  \frac{\eta_{\alpha \beta}}{2}
\Bigg( \frac{\vartheta \Big[ \begin{array}{c}  \alpha \\ \beta \end{array} \Big]}{\eta^3} \Bigg)^2
\Bigg( \frac{\vartheta \Big[ \begin{array}{c} 1/2 - \alpha \\ \beta  \end{array} \Big]}
{\vartheta \Big[ \begin{array}{c} 0 \\ 1/2 \end{array} \Big] } \Bigg)^2.
\eea
The two copies of $\hbox{Tr}({\bf 1})$ corresponds to sums over 37 and 73 states.
The twisted D3-D7 amplitudes are
\bea
\mc{A}_{37}^{(k)} & = & \int \frac{dt}{2t} \frac{1}{(2 \pi^2 t)^2} 
\left( \hbox{Tr}(\gamma^3_{\theta^k}) \hbox{Tr}(\gamma^{7\ph{g}*}_{\theta^k}) + \hbox{Tr}(\gamma^{3\ph{g}*}_{\theta^k}) 
\hbox{Tr}(\gamma^7_{\theta^k}) \right)  \nonumber \\
& & 
\Bigg[ \sum_{\alpha, \beta=0,1/2}  \frac{\eta_{\alpha \beta}}{2}
\frac{\vartheta \Big[ \begin{array}{c}  \alpha \\ \beta \end{array} \Big]}{\eta^3}
 \prod_{i=1}^{2} 
\frac{\vartheta \Big[ \begin{array}{c} 1/2 - \alpha \\ \beta + \theta_i \end{array} \Big]}
{\vartheta \Big[ \begin{array}{c} 0 \\ 1/2 + \theta_i \end{array} \Big] }
\left( - 2 \sin \pi \theta_3 \right)
\frac{\vartheta \Big[ \begin{array}{c} \alpha \\ \beta + \theta_i \end{array} \Big]}
{\vartheta \Big[ \begin{array}{c} 1/2 \\ 1/2 + \theta_i \end{array} \Big] } \Bigg].
\eea
Re-expressed in closed string channel these generate a twisted RR tadpole given by 
\be
(1_{NSNS} - 1_{RR}) 
\left( \hbox{Tr}(\gamma^3_{\theta^k}) \hbox{Tr}(\gamma^{7\ph{,}*}_{\theta^k}) + \hbox{Tr}(\gamma^{3\ph{,}*}_{\theta^k}) 
\hbox{Tr}(\gamma^7_{\theta^k}) \right)
 \int \frac{2 dl}{(2 \pi^2)^2}
\left( - 2 \sin \pi \theta_3 \right).
\ee

\subsubsection{D7-D7 amplitudes}

The final sector we may wish to consider, relevant for twisted tadpole cancellation, is the D7-D7 sector. 
The untwisted D7-D7 amplitude is
\be
\mc{A}_{77}^{0} = \int \frac{dt}{2t} Vol_4 \frac{1}{(2 \pi^2 t)^4} \hbox{Tr}(1) \sum_{\alpha, \beta=0,1/2} 
\frac{\eta_{\alpha \beta}}{2} 
\frac{\left( \vartheta \Big[ \begin{array}{c} \alpha \\ \beta \end{array} \Big] \right)^4}{\eta^{12}}.
\ee
Here $Vol_4$ represents the integral over the string centre of mass in the $4567$ directions.

The twisted D7-D7 amplitude is
\bea
\mc{A}_{77}^{(k)} & = & \int \frac{dt}{2t} \frac{(4 \sin (\pi \theta_1) \sin (\pi \theta_2))^{-2}}{(2 \pi^2 t)^2} 
\hbox{Tr}(\gamma^7_{\theta^k})  \hbox{Tr}(\gamma^{7\ph{,}*}_{\theta^k})\vert_{CP} \ti \nonumber \\
& & \sum_{\alpha, \beta=0,1/2} \frac{\eta_{\alpha \beta}}{2}  
\frac{\vartheta \Big[ \begin{array}{c} \alpha \\ \beta \end{array} \Big]}{\eta^3}
\prod_{i=1}^3 \left( - 2 \sin \pi \theta_i \right)
\frac{\vartheta \Big[ \begin{array}{c} \alpha \\ \beta + \theta_i \end{array} \Big]}
{\vartheta \Big[ \begin{array}{c} 1/2 \\ 1/2 + \theta_i \end{array} \Big] }.
\eea
The factor of $(4 \sin (\pi \theta_1) \sin (\pi \theta_2))^{-2}$ arises from integrating over the string
centre of mass in the NN directions.
Transformed to closed string channel this generates a twisted tadpole
\be
(1_{NSNS} - 1_{RR}) \frac{\hbox{Tr}(\gamma^7_{\theta}) \hbox{Tr}(\gamma^{7\ph{,}*}_{\theta})}{(4 \sin (\pi \theta_1) \sin (\pi \theta_2))^{2}}
 \int \frac{2 dl}{(2 \pi^2)^2}
\prod_{i=1}^3 \left( - 2 \sin \pi \theta_i \right).
\ee

\subsection{Magnetised Amplitudes}
\label{secMagnetised}

To compute gauge threshold corrections using the background field method 
we need the above partition functions in the presence of a background spacetime
magnetic field. This magnetic field is absent \emph{in vacuo}, and is simply a formal device to
compute the threshold corrections. 
This is turned on in the $23$ (spacetime) directions along one of the generators of the gauge group.
This shifts the open string oscillator moding in the $23$ directions.
The gauge threshold corrections can be extracted from the $\mc{O}(B^2)$ term of the expansion (\ref{vacen}).
As we are interested in threshold corrections to D3 gauge couplings we only need include the 33 and 37 amplitudes.

We briefly summarise the effect of the $B$ field on the oscillator modes \cite{acny, 9209032}. We denote the charges felt by the left and right
end of the string as $q_1$ and $q_2$, and write $\beta_1 = q_1 B, \beta_2 = q_2 B$. Neutral strings have $\beta_1 = - \beta_2$.
\begin{itemize}
\item
The oscillator moding is shifted,
\be
n^{\pm} \to n \pm \epsilon, \quad \textrm{ with } \epsilon = \frac{\tan^{-1}(\beta_1) + \tan^{-1}(\beta_2)}{\pi}.
\ee
Here $X^{\pm} = X_2 \pm iX_3$. This removes the momentum integral ($n=0$) from the partition function.
\item
The position coordinates become non-commutative, 
$$
[x^{+}, x^{-}] = \frac{\pi}{\beta_1 + \beta_2}.
$$
The integral over center of mass modes is modified,
\be
\int dx^{+} dx^{-} \to \frac{2 \pi}{[x^{+}, x^{-}]}\int dx^{+} dx^{-} = 
\frac{\beta_1 + \beta_2}{2 \pi^2} \int dx^{+} dx^{-}.
\ee
\item
The partition functions for each spin sector are modified,
\be
\quad \int dx^{+} dx^{-} \frac{1}{(2 \pi^2 t)} 
\frac{ \vartheta \Big[ \begin{array}{c} \alpha \\ \beta \end{array} \Big]}{\eta^{3}}
\longrightarrow
\frac{i(\beta_1 + \beta_2)}{2 \pi^2} 
\quad \int dx^{+} dx^{-} 
\frac{\vartheta \Big[ \begin{array}{c} \alpha \\ \beta \end{array} \Big] \left(\frac{i 
\epsilon t}{2}\right)}{\vartheta \Big[ \begin{array}{c} 1/2 \\ 1/2 \end{array} \Big] 
\left(\frac{i 
\epsilon t}{2}\right) }.
\ee
\end{itemize}
The $B$ field has no effect on the oscillator moding in the compact dimensions and the relevant
expressions are unaltered from the unmagnetised case.

\subsubsection*{Magnetised D3-D3 Amplitudes}

The magnetised untwisted D3-D3 amplitude is
\be
\mc{A}_{33}^{0} = \int \frac{dt}{2t} \frac{1}{(2 \pi^2 t)} 
\sum_{\alpha, \beta=0,1/2} \frac{\eta_{\alpha \beta}}{2}
\hbox{Tr}\left( \frac{i(\beta_1 + \beta_2)}{2 \pi^2}
\frac{\vartheta \Big[ \begin{array}{c} \alpha \\ \beta \end{array} \Big] \left(\frac{i 
\epsilon t}{2}\right)}{\vartheta \Big[ \begin{array}{c} 1/2 \\ 1/2 \end{array} \Big] 
\left(\frac{i 
\epsilon t}{2}\right) }  \right)
\frac{\left( \vartheta \Big[ \begin{array}{c} \alpha \\ \beta \end{array} \Big] \right)^3}{\eta^{9}}.
\ee
The trace is over all 33 open string states, weighted by their charges and the appropriate $\vartheta$ function.
This expression however vanishes as the untwisted D3-D3 sector preserves $\mc{N}=4$ supersymmetry and so 
cannot contribute to
gauge coupling renoralisation.

The magnetised fully twisted D3-D3 amplitudes are
\bea
\label{aaa}
\mc{A}_{33}^{(k)} & = & \int \frac{dt}{2t} \frac{1}{(2 \pi^2 t)} 
\sum_{\alpha, \beta=0,1/2}  \frac{\eta_{\alpha \beta}}{2} \nonumber \\
& & \hbox{Tr}\left( \gamma_{\theta^k} \otimes \gamma^{-1}_{\theta^k}
\frac{i(\beta_1 + \beta_2)}{2 \pi^2}
\frac{\vartheta \Big[ \begin{array}{c} \alpha \\ \beta \end{array} \Big] \left(\frac{i 
\epsilon t}{2}\right)}{\vartheta \Big[ \begin{array}{c} 1/2 \\ 1/2 \end{array} \Big] 
\left(\frac{i 
\epsilon t}{2}\right) } \right)
\prod_{i=1}^3 
\frac{\left( - 2 \sin \pi \theta_i \right) \vartheta \Big[ \begin{array}{c} \alpha \\ \beta + \theta_i \end{array} \Big]}
{\vartheta \Big[ \begin{array}{c} 1/2 \\ 1/2 + \theta_i \end{array} \Big] }.
\eea
To evaluate this we need to use the expansions (recall $\vartheta_1 = \vartheta \left[ \begin{array}{c} 1/2 \\ 1/2 \end{array} \right]$)
\bea
\vartheta_1(z) & = & 2 \pi \eta^3 z + \mc{O}(z^3), \\
\vartheta_i(z) & = & \vartheta_i(0) + \frac{z^2}{2} \vartheta_i^{''}(0) + \mc{O}(z^4), \qquad i = 2,3,4.
\eea
When expanding the $\vartheta_1$ term in the denominator, we only need consider the $\mc{O}(z)$ term as the $\mc{O}(z^3)$
term only gives an overall multiplicative prefactor to the unmagnetised partition function, which vanishes due to supersymmetry.
The $(1/2, 1/2)$ spin structure also gives a vanishing contribution: there is no $\mc{O}(B^2)$ term and the $\mc{O}(B)$ term appears
as $\sim \hbox{Tr}(\beta_1 + \beta_2) = 0$.

We can therefore simplify
\bea
\hbox{Tr}\left( \gamma_{\theta^k} \otimes \gamma^{-1}_{\theta^k}
\frac{i(\beta_1 + \beta_2)}{2 \pi^2}
\frac{\vartheta \Big[ \begin{array}{c} \alpha \\ \beta \end{array} \Big] \left(\frac{i 
\epsilon t}{2}\right)}{\vartheta \Big[ \begin{array}{c} 1/2 \\ 1/2 \end{array} \Big] 
\left(\frac{i 
\epsilon t}{2}\right) } \right) & = & - \frac{B^2 t}{16 \pi^4} \ti 
\frac{\vartheta'' \Big[ \begin{array}{c} \alpha \\ \beta \end{array} \Big]}{\eta^3}
\hbox{Tr}\Big( \left( q_L^2 + q_R^2 \right) \gamma_{\theta^k} \otimes \gamma_{\theta^k}^{-1} \Big). \nonumber
\eea

\subsubsection*{Magnetised D3-D7 Amplitudes}

The untwisted magnetised D3-D7 amplitudes are
\be
\mc{A}_{37}^{(0)} = \int \frac{dt}{2t} \frac{1}{(2 \pi^2 t)} 
\sum_{\alpha, \beta=0,1/2}  \frac{\eta_{\alpha \beta}}{2}
\hbox{Tr}\left(
\frac{i(\beta_1 + \beta_2)}{2 \pi^2}
\frac{\vartheta \Big[ \begin{array}{c} \alpha \\ \beta \end{array} \Big] \left(\frac{i 
\epsilon t}{2}\right)}{\vartheta \Big[ \begin{array}{c} 1/2 \\ 1/2 \end{array} \Big] 
\left(\frac{i 
\epsilon t}{2}\right) }  \right)
\Bigg( \frac{\vartheta \Big[ \begin{array}{c}  \alpha \\ \beta \end{array} \Big]}{\eta^3} \Bigg)
\Bigg( \frac{\vartheta \Big[ \begin{array}{c} 1/2 - \alpha \\ \beta  \end{array} \Big]}
{\vartheta \Big[ \begin{array}{c} 0 \\ 1/2 \end{array} \Big] } \Bigg)^2.
\ee
The twisted magnetised D3-D7 amplitudes are
\bea
\mc{A}_{37}^{(k)} & = & \frac{1}{3} \int \frac{dt}{2t} \frac{1}{(2 \pi^2 t)} 
\sum_{\alpha, \beta=0,1/2}  \frac{\eta_{\alpha \beta}}{2}
 \hbox{Tr}\left( \left( \gamma^3_{\theta^k} \otimes \gamma^{7\ph{g}*}_{\theta^k} + 
 \gamma^{3\ph{g}*}_{\theta^k} \otimes \gamma^7_{\theta^k} \right) \frac{i(\beta_1 + \beta_2)}{2 \pi^2}
\frac{\vartheta \Big[ \begin{array}{c} \alpha \\ \beta \end{array} \Big] \left(\frac{i 
\epsilon t}{2}\right)}{\vartheta \Big[ \begin{array}{c} 1/2 \\ 1/2 \end{array} \Big] 
\left(\frac{i 
\epsilon t}{2}\right) } \right)   \nonumber \\
& & 
\Bigg[ 
 \prod_{i=1}^{2} 
\frac{\vartheta \Big[ \begin{array}{c} 1/2 - \alpha \\ \beta + \theta_i \end{array} \Big]}
{\vartheta \Big[ \begin{array}{c} 0 \\ 1/2 + \theta_i \end{array} \Big] }
\left( - 2 \sin \pi \theta_3 \right)
\frac{\vartheta \Big[ \begin{array}{c} \alpha \\ \beta + \theta_3 \end{array} \Big]}
{\vartheta \Big[ \begin{array}{c} 1/2 \\ 1/2 + \theta_3 \end{array} \Big] } \Bigg].
\eea
We defer further evaluation of the 37 and 73 amplitudes to section \ref{37sec} when we consider a specific 
D3/D7 model.

\section{Pure D3-D3 models}

In this section we want to study threshold corrections for models of fractional D3 branes located at 
singularities. We shall focus first on abelian $\mbb{Z}_N$ singularities and subsequently on the non-Abelian
$\Delta_{27}$ singularity. Similar methods can be used to study $\mbb{Z}_N \ti \mbb{Z}_M$ singularities (cf \cite{9807082}), which we shall 
however not consider here.

\subsection{$\mbb{Z}_4$ singularity}

\begin{figure}[ht]
\label{figz4quiver}
\begin{center}
\includegraphics[width=4cm, height=4cm]{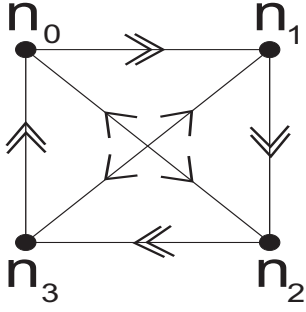}
\caption{The quiver for the $\mbb{Z}_4$ singularity.}
\end{center}
\end{figure}

The $\mbb{Z}_4$ singularity is generated by the orbifold action $\frac{1}{4}(1,1,-2)$ and the
$\mbb{Z}_4$ quiver is shown in figure \ref{figz4quiver}.
The $\beta$-function coefficient for gauge group $SU(n_0)$ is $b_0 = -3n_0 + n_1 + n_2 + n_3$.
Anomaly cancellation implies $n_1 = n_3$ and $n_0 = n_2$. If $n_0 \neq n_1$, the D3 brane
configuration by itself is anomaly free but with non-zero beta functions. 
In this case we can focus
entirely on the 33 sector.

The Chan-Paton matrix has the block-diagonal form $\Big( (\mbb{I})_{n_0}, (i\mbb{I})_{n_1}, (-\mbb{I})_{n_2},(-i I)_{n_3} \Big)$.
\bea
\hbox{Tr}\left( (q_L^2 + q_R^2) \gamma_\theta \gamma_\theta^{*} \right) & = & n_0 - n_2, \\
\hbox{Tr}\left( (q_L^2 + q_R^2) \gamma_\theta^2 \gamma_{\theta^2}^{*} \right) & = & n_0 - n_1 + n_2 - n_3.
\eea
We embed $F_{\mu \nu}$ in the $SU(n_0)$ sector, $F_{\mu \nu} = 1/2(1, -1, 0 \ldots 0)$. The contribution of $\theta$ and $\theta^3$ sectors to %%@
threshold corrections is given by 
\be
\label{def}
\frac{(n_0 - n_2)}{4} \sum_{k=1,\neq 2}^3 \int \frac{dt}{2t} \half \left( \frac{B}{2 \pi^2} \right)^2 \ti
\frac{1}{8 \pi^2} 
\sum \eta_{\alpha \beta} 
\frac{\vartheta'' \Big[ \begin{array}{c} \alpha \\ \beta \end{array} \Big]}{\eta^3 }  
\prod_{i=1}^3 \left( -2 \sin \pi \theta_i
\frac{\vartheta \Big[ \begin{array}{c} \alpha \\ \beta + \theta_i  \end{array} \Big]}
{\vartheta \Big[ \begin{array}{c} 1/2 \\ 1/2 + \theta_i \end{array} \Big] } \right).
\ee
Using the relations (\ref{ts1}) and (\ref{ts2}), we get
\bea
\sum \eta_{\alpha \beta} 
\frac{\vartheta'' \Big[ \begin{array}{c} \alpha \\ \beta \end{array} \Big]}{\eta^3 }  
\prod_{i=1}^3 \left( -2 \sin \pi \theta_i
\frac{\vartheta \Big[ \begin{array}{c} \alpha \\ \beta + \theta_i  \end{array} \Big]}
{\vartheta \Big[ \begin{array}{c} 1/2 \\ 1/2 + \theta_i \end{array} \Big] } \right)
& = & - 2 \pi \sum_{i=1}^3 
\frac{\vartheta' \Big[ \begin{array}{c} 1/2 \\ 1/2 - \theta_i  \end{array} \Big]}
{\vartheta \Big[ \begin{array}{c} 1/2 \\ 1/2 - \theta_i \end{array} \Big] } \nonumber \\
& \longrightarrow_{t \to \infty} & -2 \pi^2 \sum_i \frac{ \cos \pi \theta_i}{\sin \pi \theta_i}.
\eea
This gives for the $\theta$ + $\theta^3$ sectors
\be
\label{jc3}
\int \frac{dt}{2t} \frac{1}{4} \left( \frac{B}{2 \pi^2} \right)^2 \Big[ -2 n_0 + 2n_2 \Big].
\ee
The analogue of (\ref{def}) for the $\theta^2$ sector is $(n_0 - n_1 +n_2 -n_3) \ti$
\be
\label{jc1}
\frac{1}{4} \int \frac{dt}{2t} \half \left( \frac{B}{2 \pi^2} \right)^2 \frac{1}{8 \pi^2}
\sum \eta_{\alpha \beta} (-1)^{2 \alpha}  
\frac{\vartheta'' \Big[ \begin{array}{c} \alpha \\ \beta \end{array} \Big]}{\eta^3 }  
\frac{\vartheta \Big[ \begin{array}{c} \alpha \\ \beta \end{array} \Big]}{\eta^3 }  
\prod_{i=1}^2 \left( -2 \sin \pi \theta_i
\frac{\vartheta \Big[ \begin{array}{c} \alpha \\ \beta + \theta_i  \end{array} \Big]}
{\vartheta \Big[ \begin{array}{c} 1/2 \\ 1/2 + \theta_i \end{array} \Big] } \right).
\ee
The $(-1)^{2 \alpha}$ term arise from the action of the $2 \pi$ twist on the R sector ground state.
The above expression simplifies drastically using the identity (\ref{ts3})
\be
\sum \eta_{\alpha \beta} (-1)^{2 \alpha}  
\frac{\vartheta'' \Big[ \begin{array}{c} \alpha \\ \beta \end{array} \Big]}{\eta^3 }  
\frac{\vartheta \Big[ \begin{array}{c} \alpha \\ \beta \end{array} \Big]}{\eta^3 }  
\frac{\vartheta \Big[ \begin{array}{c} \alpha \\ \beta + \theta_1  \end{array} \Big]}
{\vartheta \Big[ \begin{array}{c} 1/2 \\ 1/2 + \theta_1 \end{array} \Big] }
\frac{\vartheta \Big[ \begin{array}{c} \alpha \\ \beta + \theta_2  \end{array} \Big]}
{\vartheta \Big[ \begin{array}{c} 1/2 \\ 1/2 + \theta_2 \end{array} \Big] } = - 4\pi^2
\ee
for $\theta_1 + \theta_2 = 1 \hbox{ mod } 2$. (\ref{jc1}) therefore becomes
\be
\label{jc2}
\int \frac{dt}{2t} \frac{1}{4} \left( \frac{B}{2 \pi^2} \right)^2 \Big[ -n_0 +n_1 - n_2 +n_3 \Big].
\ee
Note that the $\mc{N}=2$ oscillator sum collapses to a single number, and so this
is an exact expression and not merely one holding in the $t \to \infty$ limit.

Summing (\ref{jc3}) and (\ref{jc2}) to combine all sectors gives
\be
\label{threshz4}
\int \frac{dt}{2 t} \frac{1}{4} \left( \frac{B}{2 \pi^2} \right)^2 \Big[ -3n_0 + n_1+n_2 +n_3 \Big],
\ee
which is seen to reproduce the correct $\beta$ function coefficient. 

There are two comments to make here. First, 
although we formally summed over all twisted sectors, in fact the only sector that contributes
is the $\theta^2$ sector. 
The contributions of $\theta$ and $\theta^3$ sectors
are in fact seen to vanish once the anomaly cancellation conditions are imposed. 
The vanishing of the $\mc{N}=1$ $\theta$ and $\theta^3$ sectors is tied to the cancellation of fully twisted
tadpoles, which must be performed locally as these are restricted to the singularity.

Second, in the $\mc{N}=2$ $\theta^2$ sector not only do the full $\beta$ functions emerge but furthermore
the entire open string oscillator tower decoupled. Eq. (\ref{jc2}) is valid not only in the IR $t \to \infty$ limit but also
in the UV limit $t \to 0$. The decoupling can be understood as a consequence of $\mc{N}=2$ supersymmetry: only short BPS
multiplets can contribute to gauge coupling renormalisation, but all open string oscillators are non-BPS
and therefore cannot renormalise the gauge couplings. In the non-compact limit the integral (\ref{threshz4}) is therefore divergent
as $t \to 0$ with no UV cutoff. 

This divergence has a natural interpretation. The convergence of threshold corrections in the UV limit $t \to 0$ is equivalent 
to global tadpole cancellation. However in the local model there is a non-zero amplitude for the partially twisted
$\theta^2$ RR form to propagate into the bulk along the untwisted $z_3$ direction. As the charge can escape from the singularity,
this does not manifest itself as a gauge anomaly in the local D3-brane model. However in a consistent theory this tadpole 
must still be cancelled in the bulk, and the $t \to 0$ divergence reflects the fact that this global tadpole is not cancelled in the local
model.

We can understand how the cancellation will modify the threshold corrections. A consistent compact model will modify the
$\mbb{C}^3/\mbb{Z}_4$ geometry at a distance scale $R/\sqrt{\alpha'}$ from the singularity, where $R$ is the characteristic radius
of the geometry. This modification will introduce additional branes or O-planes into the model. There will then exist new states
charged under the $SU(n_0)$ gauge group, corresponding either to strings stretching between $SU(n_0)$ and bulk branes, or to winding strings
reaching round the compact space. Such state will enter the computation and modify the 
above calculation once $e^{-\pi t R^2} \sim 1$, and so this imposes
an effective cutoff $t \gtrsim 1/R^2$ on the above divergences.

\begin{figure}[ht]
\label{z4orbifold}
\begin{center}
\includegraphics[width=12cm]{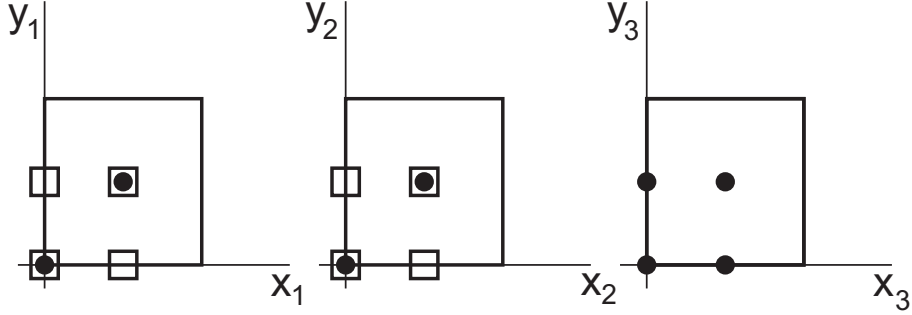}
\caption{The $T^6/\mbb{Z}_4$ orbifold. Dark circles correspond to $\theta^1$ fixed points and hollow squares correspond to
$\theta^2$ fixed points.}
\end{center}
\end{figure}

To illustrate this, we show in figure \ref{z4orbifold} the $T^6/\mbb{Z}_4$ orbifold.
As a compact space this orbifold has $h^{1,1} = 31, h^{2,1} =7$. The 31 elements of $h^{1,1}$ decomposes as 
5 untwisted 2-cycles, 16 $\theta^1$ twisted cycles stuck at the 16 $\mbb{Z}_4$ fixed points, 6 $\theta^2$ twisted cycles stuck
at $\mbb{Z}_4$ invariant combinations of $\theta^2$ fixed points, and 4 $\theta^2$ twisted cycles at $\mbb{Z}_4$ fixed points and
propagating across the third $T^2$. 

The principal point is that for each $\mbb{Z}_4$ fixed point the $\theta^2$ sector is not in homology 
uniquely associated to that fixed point: it is rather shared by the four fixed points differing by their location in the 
$(x_3, y_3)$ plane. Fractional branes at any one of these four fixed points can source an RR tadpole 
for this 2-cycle, and in a compact model we must cancel the tadpole by 
summing over the branes at all fixed points. From an open string perspective
this corresponds to including strings stretched between (for example) the (0,0,0) fixed point and the (0,0,i/2) fixed point.
Such strings would have mass $\sim R/\sqrt{\alpha'}$ and correspond to the UV cutoff on the computation of threshold corrections.

We can make this explicit. We suppose we have a stack of fractional branes on the (0,0,0) fixed point (point A) and a stack
on the (0,0,i/2) fixed point (point B). As before, we compute threshold corrections for the $SU(n_0)$ stack at point A.
There are contributions to the threshold corrections from both AA and AB strings. Following (\ref{jc2}) the AA strings give
\be
\label{jc2wind}
\int \frac{dt}{2t} \frac{1}{4} \left( \frac{B}{2 \pi^2} \right)^2 \Big[ -n_0^A +n_1^A - n_2^A +n_3^A \Big] \sum_{n,m} e^{-(n^2 + m^2) R^2 t},
\ee
where $R$ is the size of $T^3$. This incorporates the effect of AA winding strings into our previous expression (\ref{jc2}).
The global model also contains a new sector, the AB strings. These give
\be
\label{ABwind}
\int \frac{dt}{2t} \frac{1}{4} \left( \frac{B}{2 \pi^2} \right)^2 \Big[ -n_0^B +n_1^B - n_2^B +n_3^B \Big] \sum_{n,m} e^{-((n+1/2)^2 + m^2) R^2 t}.
\ee
This contribution only becomes relevant for $t \lesssim 1/R^2$. The simplest way to cancel all twisted RR tadpoles is to put
$n_0^A = n_2^A = n_1^B = n_3^B$ and $n_1^A = n_3^A = n_0^B = n_2^B$. The full global expression for the $SU(n_0)$ threshold correction is then
\be
\label{finite}
\int \frac{dt}{2t} \frac{1}{4} \left( \frac{B}{2 \pi^2} \right)^2 b_0 \left( \sum_{n,m} e^{-(n^2 + m^2) R^2 t} - \sum_{n,m} e^{-((n+1/2)^2 + m^2) R^2 %%@
t} \right).
\ee
Either through explicit numerical evaluation or through Poisson resummation into closed string channel, (\ref{finite}) is easily checked to now
be finite
in the $t \to 0$ limit, with the turn-off of the beta functions occurring at $1/t \sim (R/2)^2$ corresponding to the mass of the first AB
state.

\subsection{$\mbb{Z}_6$ singularity}

The $\mbb{Z}_6$ singularity is generated from the orbifold action $\frac{1}{6}(1,1,-2)$. The $\theta, \theta^2,
\theta^4$ and $\theta^5$ sectors are all $\mc{N}=1$ sectors with $\theta^3$ the only 
$\mc{N}=2$ sector. The quiver for $\mbb{Z}_6$ is shown in
figure \ref{figz6quiver} below.
\begin{figure}[ht]
\label{figz6quiver}
\begin{center}
\includegraphics[width=4cm, height=4cm]{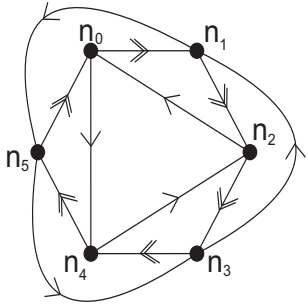}
\caption{The quiver for the $\mbb{Z}_6$ singularity.}
\end{center}
\end{figure}
The $\beta$ function coefficient for $SU(n_0)$ is $b_0 = -3n_0 + (n_1 + n_5) + \frac{n_2 + n_4}{2}$. 
Anomaly cancellation implies
$n_2 + 2n_5 = 2n_1 + n_4$ plus cyclic permutations. 
The anomaly cancellation conditions constrain $n_0 = n_2 = n_4$ and $n_1 = n_3 = n_5$, leaving two
independent parameters.

Using the block-diagonal Chan-Paton matrix 
$\gamma_{\theta} = \Big( (\mbb{I})_{n_0}, (e^{\frac{\pi i}{3}} \mbb{I})_{n_1}, 
(e^{\frac{2 \pi i}{3}} \mbb{I})_{n_2}, \ldots , (e^{\frac{5 \pi i}{3}}\mbb{I})_{n_5} \Big),$
we have
\bea
\label{sixtr}
\hbox{Tr}\left( (q_L^2 + q_R^2) \gamma_\theta \gamma_\theta^{*} \right) & = & (n_0 - n_3) + \frac{n_1 - n_4}{2} - \frac{n_2 - n_5}{2}, \\
\hbox{Tr}\left( (q_L^2 + q_R^2) \gamma_\theta^2 \gamma_{\theta^2}^{*} \right) & = & (n_0 + n_3) - \frac{n_1 + n_4}{2} - \frac{n_2 + n_5}{2}, \\
\hbox{Tr}\left( (q_L^2 + q_R^2) \gamma_\theta^3 \gamma_{\theta^3}^{*} \right) & = & n_0 - n_1 + n_2 - n_3 +n_4 - n_5.
\eea
No difficulties are encountered in the computation of threshold corrections. Summing all $\mc{N}=1$ sectors gives
in the $t \to \infty$ limit:
\be
\label{sixn1}
\int \frac{dt}{2t} \frac{1}{4} \left( \frac{B}{2 \pi^2} \right)^2 \frac{-1}{12} \Big[ 5(2n_0 + n_1 -n_2 -2n_3 -n_4 +n_5)
+9(2n_0 -n_1 -n_2 +2n_3 -n_4 -n_5) \Big].
\ee
The $\theta^3$ $\mc{N}=2$ sector gives the exact result
\be
\label{sixn2}
\int \frac{dt}{2t} \frac{1}{4} \left( \frac{B}{2 \pi^2} \right)^2 \frac{-8}{12} \Big[ n_0 - n_1 + n_2 -n3 + n_4  -n_5 \Big].
\ee
Combining (\ref{sixn1}) and (\ref{sixn2}) we obtain
\be
\int \frac{dt}{2t} \frac{1}{4} \left( \frac{B}{2 \pi^2} \right)^2 \Big[ -3n_0 +n_1 + \frac{n_2 + n_4}{2} + n_5 \Big],
\ee
with the correct IR $\beta$ function coefficient.

The same physics that occurred in the $\mbb{Z}_4$ case recurs here. Although the threshold corrections are a formal sum over all sectors,
the contribution (\ref{sixn1}) of the $\mc{N}=1$ sectors vanishes when anomaly
cancellation is imposed. This is due to the fact that the $\theta$ and $\theta^2$ sectors are associated to twisted RR states 
that are tied to the singularity, and so the resulting RR tadpole must be cancelled locally.
In contrast, the single $\mc{N}=2$ sector need not have its RR tadpole cancelled locally, allowing the existence of
non-zero beta functions.

\subsection{$\mbb{Z}_6'$ singularity}

\begin{figure}[ht]
\label{figz6primequiver}
\begin{center}
\includegraphics[width=4cm, height=4cm]{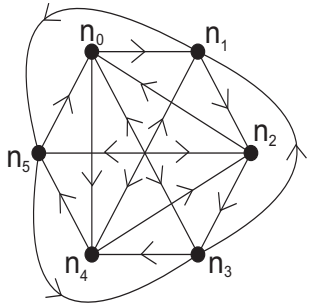}
\caption{The quiver for the $\mbb{Z}_6'$ singularity.}
\end{center}
\end{figure}

The $\mbb{Z}_6'$ singularity is generated by the action $\frac{1}{6}(1,2,-3)$. The principle difference to
 $\mbb{Z}_4$ or $\mbb{Z}_6$ is that $\mbb{Z}_6'$ has multiple $\mc{N}=2$ sectors. The quiver is shown in figure \ref{figz6primequiver}.
The $\beta$-function coefficient for $SU(n_0)$ is $b_0 = -3n_0 + \frac{n_1}{2} + \frac{n_2}{2} + n_3 + \frac{n_4}{2} + \frac{n_5}{2}$.
The anomaly cancellation conditions are the cyclic permutations of $n_0 + n_1 = n_3 + n_4$. The presence of 3 $\mc{N}=2$ sectors
implies that the solutions to the anomaly conditions have (3+1) free parameters. The solutions are
$$
n_0 = n_2 + n_3 - n_5, \qquad n_1 = -n_2 + n_4 + n_5.
$$
The Chan-Paton traces take the same values as for the $\mbb{Z}_6$ case (\ref{sixtr}).  
The threshold corrections are then easily computed, and the various sectors give
\bea
\theta + \theta^5: & \qquad & \int \frac{dt}{2t} \frac{1}{4} \left( \frac{B}{2 \pi^2} \right)^2 \frac{-8}{12} \Big[
(2n_0 + n_1 -n_2 -2n_3 -n_4 +n_5) \Big], \\
\theta^2 + \theta^4:  & \qquad & \int \frac{dt}{2t} \frac{1}{4} \left( \frac{B}{2 \pi^2} \right)^2 \frac{-6}{12} \Big[
(2n_0 - n_1 -n_2 +2n_3 -n_4 -n_5) \Big], \\
\theta^3: & \qquad & 
\int \frac{dt}{2t} \frac{1}{4} \left( \frac{B}{2 \pi^2} \right)^2 \frac{-8}{12} \Big[
(n_0 - n_1 +n_2 -n_3 +n_4 -n_5) \Big].
\eea
Combining all sectors we obtain the correct $SU(n_0)$ $\beta$-function
\be
\int \frac{dt}{2t} \frac{1}{4} \left( \frac{B}{2 \pi^2} \right)^2 \Big[ -3n_0 + \frac{n_1 + n_2}{2} +n_3 + \frac{n_4 +n_5}{2} \Big].
\ee
The same physics is again evident: the $\mc{N}=1$ sectors ($\theta$ and $\theta^5$) vanish
once anomaly cancellation is imposed. The threshold corrections 
are sourced entirely by the $\mc{N}=2$ sectors, for which the oscillator sum reduces to a single constant.

\subsection{$\Delta_{27}$ singularity}

We finally consider the $\Delta_{27}$ singularity. Unlike the previous cases, this is a nonabelian singularity, studied
for example in \cite{9811183, 9811258, 0105042}.
This also exists as a particular case of the $dP_8$ singularity \cite{0508089}, which is the most
general of the del Pezzo family. The $\Delta_{27}$ quiver is  
\begin{figure}[ht]
\begin{center}
\includegraphics[width=7cm]{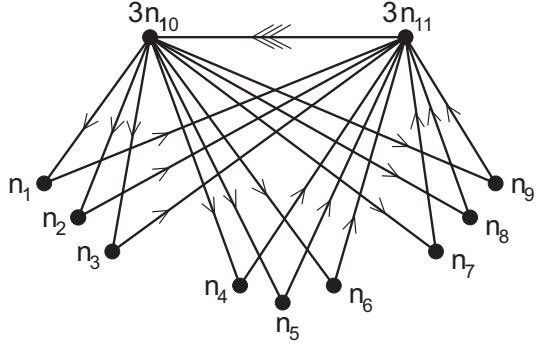}
\caption{The quiver for the $\Delta_{27}$ singularity.}
\end{center}
\label{Delta27quiver}
\end{figure}

The anomaly cancellation conditions are
\be
\label{d27anom}
n_{10} = n_{11} = \frac{(n_1 + n_2 + n_3 + n_4 + n_5 + n_6 + n_7 + n_8 + n_9)}{9}.
\ee
We start by reviewing the basic properties of $\Delta_{27}$. $\Delta_{27}$ is the non-abelian finite subgroup
of $SU(3)$ generated by
\bea
e_1: (z_1, z_2, z_3) & \to & (\omega z_1, \omega^2 z_2, z_3), \nonumber \\
e_2: (z_1, z_2, z_3) & \to & (z_1, \omega z_2, \omega^2 z_3), \nonumber \\
e_3: (z_1, z_2, z_3) & \to & (z_3, z_1, z_2).
\eea
The generators satisfy 
\be
\label{genrep}
e_1^3 = e_2^3 = e_3^3 =1, \quad e_1 e_2 = e_2 e_1, \quad e_3 e_1 = e_2 e_3, \quad e_3 e_2 = e_1^2 e_2^2 e_3.
\ee
The 27 elements of the group can be written as $e_1^{\alpha} e_2^{\beta} e_3^{\gamma}$, with
$\alpha, \beta, \gamma = 1 \ldots 3$. The eleven conjugacy classes are
$$
\{1\}, \{e_1 e_2^2\}, \{e_1^2, e_2 \}, \{e_1, e_2, e_1^2 e_2^2 \}, \{e_1^2, e_2^2, e_1 e_2 \},
\{e_2 e_3, e_1 e_3, e_1^2 e_2^2 e_3 \}, \{ e_1 e_2 e_3, e_1^2 e_3, e_2^2 e_3 \},$$
$$
\{ e_3^2, e_1 e_2^2 e_3^2, e_1^2 e_2 e_3^2 \}, \{ e_1 e_3^2, e_1^2 e_2^2 e_3^2, e_2 e_3^2 \},
\{ e_1^2 e_3^2, e_2^2 e_3^2, e_1 e_2 e_3^2 \}, \{ e_3^2, e_1 e_2^2 e_3^2, e_1^2 e_2 e_3^2\}.
$$
Corresponding to these are eleven irreducible representations, ${\bf{3}} + {\bf{3}}^{*} + 9 \ti {\bf{1}}$.
Using the relations (\ref{genrep}) it is easy to see that the nine 1-dimensional irreps are given by
\be
\gamma_{e1} = \gamma_{e2} = \omega^{\alpha}, \gamma_{e3} = \omega^{\beta},
\ee
with $\alpha, \beta = 0,1,2$. The 3-dimensional irreps are given by the defining representation and its complex
conjugate,
\bea
\gamma^{\bf{3}}_{e_1} = \left( \begin{array}{ccc} \omega & 0 & 0 \\ 0 & \omega^2 & 0 \\ 0 & 0 & 1 \end{array} \right),
\gamma^{\bf{3}}_{e_2} = \left( \begin{array}{ccc} 1 & 0 & 0 \\ 0 & \omega & 0 \\ 0 & 0 & \omega^2 \end{array} \right),
\gamma^{\bf{3}}_{e_3} = \left( \begin{array}{ccc} 0 & 0 & 1 \\ 1 & 0 & 0 \\ 0 & 1 & 0 \end{array} \right), & & \\
\gamma^{\bf{3}^{*}}_{e_1} = \left( \begin{array}{ccc} \omega^2 & 0 & 0 \\ 0 & \omega & 0 \\ 0 & 0 & 1 \end{array} \right),
\gamma^{\bf{3}^{*}}_{e_2} = \left( \begin{array}{ccc} 1 & 0 & 0 \\ 0 & \omega^2 & 0 \\ 0 & 0 & \omega \end{array} \right),
\gamma^{\bf{3}^{*}}_{e_3} = \left( \begin{array}{ccc} 0 & 0 & 1 \\ 1 & 0 & 0 \\ 0 & 1 & 0 \end{array} \right). & & 
\eea
The regular representation decomposes as $27 = (3 \ti {\bf 3}) \oplus (3 \ti {\bf 3}) \oplus {\bf 1}
\oplus {\bf 1} \oplus {\bf 1} \oplus {\bf 1} \oplus {\bf 1} \oplus {\bf 1} \oplus {\bf 1} \oplus {\bf 1} \oplus {\bf 1}$.
Using the Chan-Paton matrices for the regular representation it is straightforward to derive the $\Delta_{27}$ quiver shown in figure %%@
\ref{Delta27quiver}.

The partition function is given by
$$
\hbox{Tr} \left[ \frac{(1 + e_1 + e_1^2)}{3} \frac{(1 + e_2 + e_2^2)}{3} \frac{(1 + e_3 + e_3^2)}{3} e^{-H t} \right].
$$
To evaluate this we first evaluate the action of each group element on the oscillator tower. For explicitness consider the
states 
\be
\label{sta}
\psi^{z_1}_{-\lambda} |0\rangle, \psi^{z_2}_{-\lambda} |0\rangle, \psi^{z_3}_{-\lambda} |0\rangle.
\ee 
For an element
of the form $e_1^{\alpha} e_2^{\beta}$ the above states are eigenstates of the group element. However for elements of the form
$e_1^{\alpha} e_2^{\beta} e_3^{\gamma}$ this no longer holds. For example, the eigenstates of $e_3$ are 
$$
\psi^{z_1}_{-\lambda} |0\rangle +  \psi^{z_2}_{-\lambda} |0\rangle + \psi^{z_3}_{-\lambda} |0\rangle, \quad 
\psi^{z_1}_{-\lambda} |0\rangle +  \omega \psi^{z_2}_{-\lambda} |0\rangle + \omega^2 \psi^{z_3}_{-\lambda} |0\rangle, \quad 
\psi^{z_1}_{-\lambda} |0\rangle +  \omega^2 \psi^{z_2}_{-\lambda} |0\rangle + \omega \psi^{z_3}_{-\lambda} |0\rangle.
$$
The action of the varying group elements on the states (\ref{sta}) then gives
\bea
1: & & (1+q)(1+q)(1+q), \nonumber \\
e_1 e_2^2: & & (1+\omega q)(1+ \omega q) (1 + \omega q), \nonumber \\
e_2 e_1^2: & & (1+\omega^2 q)(1+ \omega^2 q) (1 + \omega^2 q), \nonumber \\
\hbox{all other group elements}: & & (1 + q)(1 + \omega q)(1 + \omega^2 q).
\eea
The oscillator towers for each group element are therefore (with $\sum \theta_i = 0$) given by
\bea
1: & & \Bigg( \frac{\vartheta \Big[ \begin{array}{c} \alpha \\ \beta \end{array} \Big]}{\eta^3 } \Bigg)^4 , \nonumber \\
e_1 e_2^2: & & \Bigg( \frac{\vartheta \Big[ \begin{array}{c} \alpha \\ \beta \end{array} \Big]}{\eta^3 } \Bigg)
\prod_{i=1}^3 \left( -2 \sin \pi \theta_i
\frac{\vartheta \Big[ \begin{array}{c} \alpha \\ \beta + \theta_i  \end{array} \Big]}
{\vartheta \Big[ \begin{array}{c} 1/2 \\ 1/2 + \theta_i \end{array} \Big] } \right)
, \nonumber \\
e_2 e_1^2: & & \Bigg( \frac{\vartheta \Big[ \begin{array}{c} \alpha \\ \beta \end{array} \Big]}{\eta^3 } \Bigg)
\prod_{i=1}^3 \left( -2 \sin \pi \theta_i
\frac{\vartheta \Big[ \begin{array}{c} \alpha \\ \beta + \theta_i  \end{array} \Big]}
{\vartheta \Big[ \begin{array}{c} 1/2 \\ 1/2 + \theta_i \end{array} \Big] } \right)
, \nonumber \\
\hbox{all other group elements}: & & \Bigg( \frac{\vartheta \Big[ \begin{array}{c} \alpha \\ \beta \end{array} \Big]}{\eta^3 } \Bigg)^2
\prod_{i=1}^2 \left( -2 \sin \pi \theta_i
\frac{\vartheta \Big[ \begin{array}{c} \alpha \\ \beta + \theta_i  \end{array} \Big]}
{\vartheta \Big[ \begin{array}{c} 1/2 \\ 1/2 + \theta_i \end{array} \Big] } \right).
\eea
This corresponds to one $\mc{N}=4$ sector, twenty-four $\mc{N}=2$ sectors and two $\mc{N}=1$ sectors. From here the procedure to 
compute threshold corrections is very similar to the abelian orbifolds above: we compute the correction in each
individual twisted sector and sum over all sectors. Twisted tadpole cancellation requires that the Chan-Paton traces for the
two $\mc{N}=1$ sectors vanish. Now,
\bea
\label{abf}
\hbox{Tr}(\gamma_{e_1^2 e_2}) & = & (n_1 + n_2 + n_3 + n_4 + n_5 + n_6 + n_7 + n_8 + n_9) + 9 \omega^2 n_{10} + 9 \omega n_{11}, \nonumber \\
\hbox{Tr}(\gamma_{e_1^2 e_2}) & = & (n_1 + n_2 + n_3 + n_4 + n_5 + n_6 + n_7 + n_8 + n_9) + 9 \omega n_{10} + 9 \omega^2 n_{11}.
\eea
The vanishing of twisted tadpoles therefore requires
\be
n_{10} = n_{11} = \frac{(n_1 + n_2 + n_3 + n_4 + n_5 + n_6 + n_7 + n_8 + n_9)}{9}.
\ee
This reproduces the anomaly cancellation conditions (\ref{d27anom}).

From the $\Delta_{27}$ quiver it is easy to see that only the $SU(n_1) \to SU(n_9)$ gauge groups can have
non-zero beta functions. We focus on the $SU(n_1)$ case for which the 1-dimensional representation is trivial.
Using the same formalism and results as for the abelian orbifolds, we find that
the annulus amplitude from the $\mc{N}=1$ sectors is
\be
\frac{1}{27} \int \frac{dt}{2t} \half \left( \frac{B}{2 \pi^2} \right)^2 \frac{-2 \pi^2}{8 \pi^2}
\ti 9 \left[ \sum_{\mc{N}=1} \half \left[ \hbox{Tr}(\gamma_{\theta}) + \hbox{Tr}(\gamma_\theta^{*}) \right] \right].
\ee
Using (\ref{abf}), we obtain
$$
\sum_{\mc{N}=1} \half \left[ \hbox{Tr}(\gamma_{\theta}) + \hbox{Tr}(\gamma_\theta^{*}) \right] = 2 \left(
n_1 + n_2 + \ldots + n_9 \right) - 9n_{10} - 9n_{11},
$$
giving an overall contribution from $\mc{N}=1$ sectors of
\be
\int \frac{dt}{2t} \frac{1}{4} \left( \frac{B}{2 \pi^2} \right)^2 \left[ - \frac{1}{3}\left( n_1 + n_2 + \ldots + n_9 \right)
+ \frac{3}{2} \left( n_{10} + n_{11} \right) \right].
\ee
Of course this vanishes when anomaly cancellation is imposed.

The $\mc{N}=2$ sectors give a contribution
\be
\frac{1}{27} \int \frac{dt}{2t} \half \left( \frac{B}{2 \pi^2} \right)^2 \frac{-4 \pi^2}{8 \pi^2}
\ti 3 \left[ \sum_{\mc{N}=2} \half \left[ \hbox{Tr}(\gamma_{\theta}) + \hbox{Tr}(\gamma_\theta^{*}) \right] \right].
\ee
In this case
$$
\sum_{\mc{N}=2} \half \left[ \hbox{Tr}(\gamma_{\theta}) + \hbox{Tr}(\gamma_\theta^{*}) \right] = 3 \left(
8n_1 - n_2 - n_3 - \ldots - n_9) \right)
$$
with the result that $\mc{N}=2$ sectors contribute
\be
\int \frac{dt}{2t} \frac{1}{4} \left( \frac{B}{2 \pi^2} \right)^2 \left[ -3n_1 + \frac{1}{3}\left( n_1 + n_2 + \ldots + n_9 \right)\right].
\ee
Combining both $\mc{N}=1$ and $\mc{N}=2$ sectors, we obtain
\be
\int \frac{dt}{2t} \frac{1}{4} \left( \frac{B}{2 \pi^2} \right)^2 \left[ -3n_1 + \frac{3}{2} (n_{10} + n_{11}) \right],
\ee
which is precisely the $\beta$-function of the $SU(n_1)$ gauge theory.

The exact same physics has recurred here as for the simpler examples of Abelian orbifolds. The $\beta$-functions come entirely
from the $\mc{N}=2$ sector, for which the string tower decouples. From a closed string perspective, the $\beta$-function
corresponds to the propagation of a twisted RR state into the bulk. This is a tadpole and is necessarily divergent
when analysed from a purely local perspective. 

We can be more precise and relate this to the geometry of the $\Delta_{27}$ singularity. The $\Delta_{27}$ singularity
is part of the moduli space of the del Pezzo 8 ($dP_8$) singularity. The geometry of a $dP_n$ consists of one 4-cycle and
$(n+1)$ 2-cycles. There is one collapsing 4-cycle at the singularity and $n+1$ collapsing 2-cycles. One 2-cycle is dual to
the 4-cycle: both the cycle and its dual are compact in the non-compact local model. These cycles represent 
the $\mc{N}=1$ sectors.\footnote{$dP_n$ corresponds to $dP_0 \equiv \mbb{P}^2$ with $n$ points blown up into $\mbb{P}^1$s.
The $dP_0 \equiv \mbb{P}^2 \equiv \mbb{C}^3/\mbb{Z}_3$ singularity has one 4-cycle and one 2-cycle, which are dual to each other.
These correspond to the $\mc{N}=1$ twisted sectors of the orbifold. The $\mc{N}=1$ twisted sectors of $\Delta_{27}$ are, in a sense,
inherited from the $\mc{N}=1$ twisted sectors of $\mbb{C}^3/\mbb{Z}_3$.} The RR forms on these cannot propagate into the bulk and so
tadpole cancellation must take place locally for a consistent theory. This manifests itself as the anomaly cancellation condition that
fixes $n_{10}$ and $n_{11}$.

The other $n$ 2-cycles have dual 4-cycles that are non-compact. At the level of the local model it is simply unknown whether
these 2-cycles represent 2-cycles of the Calabi-Yau, or merely 2-cycles 
of the del Pezzo.\footnote{For example, \cite{9603161} gives a case of a $dP_8$ singularity where only one 2-cycle of the $dP_8$
is non-trivial in the Calabi-Yau. A variety of examples of this phenomenon are also given in the recent paper
 \cite{08053361}.}
Furthermore, at the level of the local model it is
also unknown whether or not there are other 2-cycles in the Calabi-Yau that are in the same homology class but are spatially separated
from the $dP_8$ singularity.\footnote{An analogy here is the resolved conifold: the 2-cycle present in the resolved conifold 
geometry likewise need not be the unique representative of its homology class. I thank Xenia de la Ossa for discussions on this point.}
 As a result there can be no purely local consistency condition from these cycles. This manifests itself
in the high level of freedom in satisfying the anomaly cancellation conditions: we cannot restrict beyond $\sum_{i=1}^9 n_i = 9n_{10} = 9n_{11}$.
The eight degrees of freedom in solving $\sum_{i=1}^9 n_i = 9n_{10} = 9n_{11}$ (once the overall $N$ is fixed), giving eight
independent beta functions, correspond in closed string channel to sources for RR forms along these 8 cycles.

\begin{figure}[ht]
\label{globalworldsheet}
\begin{center}
\includegraphics[width=8cm, height=6cm]{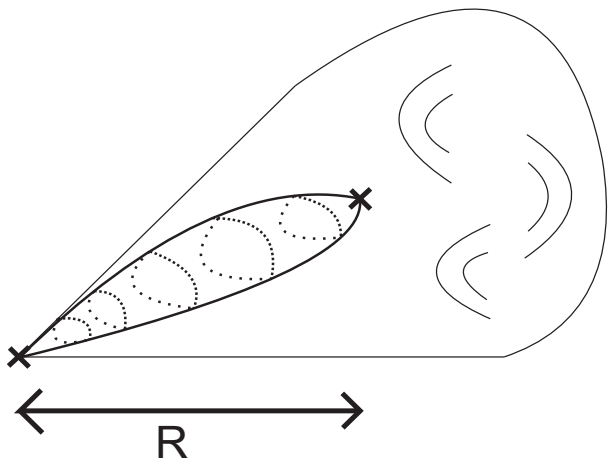}
\caption{An illustration of a worldsheet configuration that is missed by a purely local computation of threshold corrections.
From an open string perspective this corresponds to a string of mass $R/\sqrt{\alpha'}$ stretching from one 
fractional brane stack to another, whereas
in closed string language this represent sources and sinks of twisted RR charge.}
\end{center}
\end{figure}

While these tadpoles are consistent in a local model, in a global model they must necessarily vanish. This can occur either because several of these
2-cycles are homologically identical, or if there exist other distant representatives in the same homology class that also source the same
RR tadpole. Either way, the global cancellation of tadpoles requires 
knowledge of the structure of the compact space that is hidden from the local model. The precise 
nature of this is model-dependent. However what is model-independent is that in the large-radius limit 
it appears at a distance $\sim R (2 \pi \sqrt{\alpha'})$ from the singularity, where $R$ is the Calabi-Yau radius.
From the open string point of view, this corresponds to the existence of new states, of approximate mass $\frac{R}{\sqrt{\alpha'}}$,
that are not present in the local model (for example strings stretching from one brane stack to another, or winding around the compact space
back to the singularity). As described for the $T^6/\mbb{Z}_4$ case, such 
states must be included in the computation of threshold corrections for $t \lesssim \frac{1}{R^2}$
(when $e^{-Ht} \sim 1$). This is illustrated in figure \ref{globalworldsheet}. 
As stringy consistency requires that the resulting expression is finite, we can incorporate the effect of a
global compact embedding by cutting the integral off at $t \sim \frac{1}{R^2}$, which corresponding to running from a scale $\sim R M_s$.

\subsection{Comparison with Field Theory}

In all cases the above string calculations agree with the field theory result. The low energy couplings unify
at a scale $M_X = R M_s$ rather than the naive scale of $M_s$. From a string point of view, the scale $R M_s$ arises
because, while the gauge groups may be local, tadpole cancellation is not. The fractional brane configurations source
a closed string RR tadpole that through open/closed string duality is precisely equivalent to the running gauge couplings.
This tadpole is necessarily divergent in the purely local model, which does not know whether it has been consistently
embedded in a global background. Heuristically, the appearance of the scale $M_X = R M_s$ 
can be understood from the need to reach the bulk to know whether the tadpole has been cancelled or not.

From a purely open string perspective, the fact that the beta functions arise only from $\mc{N}=2$ sectors
means that the oscillator sum reduces to a constant: only BPS states can renormalise the gauge couplings, and 
open string excitations are non-BPS. The beta functions therefore do not see the string scale as a threshold but instead
continute evolving beyond it.

A similar unification at a super-stringy scale $M_X = R M_s$ was found for certain orientifold models in \cite{9906039}.
It would be interesting to see whether the underlying physics is similar. However direct comparison is not
straightforward as we have considered D3 branes whereas that paper considered the gauge couplings 
on D9 branes in globally consistent D5/D9 models.   

\section{Models involving both D3 and D7 branes}
\label{37sec}

We next consider local models involving both D3 and D7 branes, where the D7 wraps both a bulk and a collapsed cycle.
We will focus on models based around the $\mbb{C}^3/\mbb{Z}_3$ singularity with twist 
vector $\frac{1}{3}(1,1,-2)$ such as considered in \cite{aiqu} (see also \cite{08105560}). The quiver for these models
is shown in figure 7.
\begin{figure}[ht]
\label{delPezz0}
\begin{center}
\includegraphics[width=6cm]{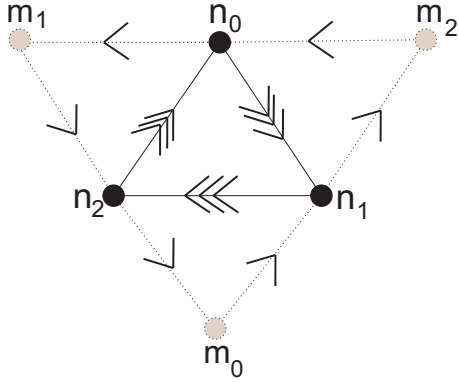}
\caption{The quiver for D3/D7 branes at the $\mbb{C}^3/\mbb{Z}_3$ singularity. The dark circles, labelled $n_i$,
represent D3 gauge groups and the grey circles, labelled $m_i$, represent D7 gauge groups.}
\end{center}
\end{figure}

Anomaly cancellation requires $3n_i + m_i = 3n_j + m_j$ for all $i,j$, giving
$3n_0 + m_0 = 3(n_1 + n_2)/2 + (m_1 + m_2)/2$.

\subsection{Tadpole cancellation}

In section \ref{subsecunmag} we wrote down the tadpoles originating from 33, 37, and 77 sectors.
The summation of the annulus amplitude over all these sectors generates a twisted closed string divergence
in the $l \to \infty$ limit, given by
\be
\mc{A}^{(k)} = (1_{NSNS} - 1_{RR}) \left\vert \prod_{i=1}^3 \left( - 2 \sin \pi \theta_i \right) \hbox{Tr}(\gamma^3_{\theta}) 
+ (- 2 \sin \pi \theta_3) \hbox{Tr}(\gamma^7_{\theta}) \right\vert^2
\int \frac{2 dl}{(2 \pi^2)^2}
\ee
The tadpole cancellation conditions are therefore
\be
\prod_{i=1}^3 \left( - 2 \sin \pi \theta_i \right) \hbox{Tr}(\gamma^3_{\theta}) 
+ (- 2 \sin \pi \theta_3) \hbox{Tr}(\gamma^7_{\theta}) = 0.
\ee
These can be checked to be equivalent to the anomaly cancellation conditions for the quiver field theory \cite{aiqu}.

There are in addition untwisted tadpoles that arise and do not vanish - for example there is 
the tadpole due to 
D3 charge and also that from the bulk cycle on which the 7-brane is wrapped.
In a fully consistent theory 
the untwisted tadpoles will be cancelled by bulk O3/O7 planes, contributing to
Mobius Strip (MS) or Klein Bottle (KB) diagrams. However no orientifolds are present in the local model
and so we again restrict ourselves to only the annulus diagrams.

\subsection{33 Amplitudes}

We first write down the magnetised 33 amplitdues following section (\ref{secMagnetised}) and eq. (\ref{aaa}). 
The trace over charged states in (\ref{aaa})
simplifies to
\bea
\hbox{Tr}\left( \gamma_{\theta^k} \otimes \gamma^{-1}_{\theta^k}
\frac{i(\beta_1 + \beta_2)}{2 \pi^2}
\frac{\vartheta \Big[ \begin{array}{c} \alpha \\ \beta \end{array} \Big] \left(\frac{i 
\epsilon t}{2}\right)}{\vartheta \Big[ \begin{array}{c} 1/2 \\ 1/2 \end{array} \Big] 
\left(\frac{i 
\epsilon t}{2}\right) } \right)& = & 
- \frac{B^2 t}{16 \pi^4} \ti 
\frac{\vartheta'' \Big[ \begin{array}{c} \alpha \\ \beta \end{array} \Big]}{\eta^3}
\Big( n_0 - \frac{(n_1 + n_2)}{2} \Big).
\eea
This expression holds equally for the $\theta$ and $\theta^2$ twists.

The expression for the threshold corrections can be further simplified using the identity (\ref{ts1}), 
putting the partition function in the form
\be
\label{aab}
\frac{1}{3} \sum_{k=1}^2 \int \frac{dt}{2t } \half \left(\frac{B}{2 \pi^2}\right)^2 \frac{-2 \pi}{8 \pi^2} \Big[ n_0 - \frac{n_1 + n_2}{2} \Big]
\Bigg( \sum_{i=1}^3 \frac{\vartheta' \Big[ \begin{array}{c} 1/2 \\ 1/2 - \theta_i \end{array} \Big]}
{\vartheta \Big[ \begin{array}{c} 1/2 \\ 1/2 - \theta_i \end{array} \Big] } \Bigg)  \left( \prod_i -2 \sin( \pi \theta_i) \right).
\ee
The IR limit of (\ref{aab}) can be evaluated using the identity (\ref{ts2}), giving
\be
\label{33aiqu}
\left( -\int \frac{dt}{2t} \frac{1}{4} \left( \frac{B}{2 \pi^2} \right)^2 \right) \ti \left(-3n_0 + \frac{3(n_1 + n_2)}{2}\right).
\ee
This reproduces the contribution to the $\beta$-function from D3-D3 states.

\subsubsection{D3-D7}

The untwisted magnetised D3-D7 amplitudes are
\be
\mc{A}_{37}^{(0)} = \int \frac{dt}{2t} \frac{1}{(2 \pi^2 t)} 
\sum_{\alpha, \beta=0,1/2}  \frac{\eta_{\alpha \beta}}{2}
\hbox{Tr}\left(
\frac{i(\beta_1 + \beta_2)}{2 \pi^2}
\frac{\vartheta \Big[ \begin{array}{c} \alpha \\ \beta \end{array} \Big] \left(\frac{i 
\epsilon t}{2}\right)}{\vartheta \Big[ \begin{array}{c} 1/2 \\ 1/2 \end{array} \Big] 
\left(\frac{i 
\epsilon t}{2}\right) }  \right)
\Bigg( \frac{\vartheta \Big[ \begin{array}{c}  \alpha \\ \beta \end{array} \Big]}{\eta^3} \Bigg)
\Bigg( \frac{\vartheta \Big[ \begin{array}{c} 1/2 - \alpha \\ \beta  \end{array} \Big]}
{\vartheta \Big[ \begin{array}{c} 0 \\ 1/2 \end{array} \Big] } \Bigg)^2.
\ee
The trace is over all D3-D7 and D7-D3 states. Unlike for D3-D3 strings, the untwisted D3-D7 sector preserves $\mc{N}=2$ 
supersymmetry and can therefore contribute to gauge coupling renormalisation.
In this case
\bea
\hbox{Tr}\left(
\frac{i(\beta_1 + \beta_2)}{2 \pi^2}
\frac{\vartheta \Big[ \begin{array}{c} \alpha \\ \beta \end{array} \Big] \left(\frac{i 
\epsilon t}{2}\right)}{\vartheta \Big[ \begin{array}{c} 1/2 \\ 1/2 \end{array} \Big] 
\left(\frac{i 
\epsilon t}{2}\right) } \right) 
& = & 
- \frac{B^2 t}{16 \pi^4} \ti 
\frac{\vartheta'' \Big[ \begin{array}{c} \alpha \\ \beta \end{array} \Big]}{\eta^3}
\Big( n_0^7 + n_1^7 + n_2^7 \Big).
\eea
The resulting expression for threshold corrections is
\be
\frac{1}{3} \int \frac{dt}{2t} \left(\frac{B}{2 \pi^2} \right)^2 \frac{1}{8 \pi^2} \Big[ n_0^7 + n_1^7 + n_2^7 \Big]
\sum \frac{\eta_{\alpha \beta}}{2}
\frac{\vartheta'' \Big[ \begin{array}{c} \alpha \\ \beta \end{array} \Big]}{\eta^3 }  
\Bigg( \frac{\vartheta \Big[ \begin{array}{c}  \alpha \\ \beta \end{array} \Big]}{\eta^3} \Bigg)
\Bigg( \frac{\vartheta \Big[ \begin{array}{c} 1/2 - \alpha \\ \beta  \end{array} \Big]}
{\vartheta \Big[ \begin{array}{c} 0 \\ 1/2 \end{array} \Big] } \Bigg)^2.
\ee
Now,
\be
\sum \frac{\eta_{\alpha \beta}}{2}
\frac{\vartheta'' \Big[ \begin{array}{c} \alpha \\ \beta \end{array} \Big]}{\eta^3 }  
\Bigg( \frac{\vartheta \Big[ \begin{array}{c}  \alpha \\ \beta \end{array} \Big]}{\eta^3} \Bigg)
\Bigg( \frac{\vartheta \Big[ \begin{array}{c} 1/2 - \alpha \\ \beta  \end{array} \Big]}
{\vartheta \Big[ \begin{array}{c} 0 \\ 1/2 \end{array} \Big] } \Bigg)^2 = 2\pi^2,
\ee
and so the whole oscillator sum reduces to a single number. We obtain
\be
\label{d3d7betauntwisted}
\left( -\int \frac{dt}{2t} \frac{1}{4} \left( \frac{B}{2 \pi^2} \right)^2 \right) \ti \frac{\left(n_0^7 + n_1^7 + n_2^7 \right)}{3}.
\ee
This represents the full contribution of the untwisted sector to the threshold corrections; as mentioned above there is no oscillator sum.

The twisted magnetised D3-D7 amplitudes are
\bea
\label{gag}
\mc{A}_{37}^{(k)} & = & \frac{1}{3} \int \frac{dt}{2t} \frac{1}{(2 \pi^2 t)} 
\sum_{\alpha, \beta=0,1/2}  \frac{\eta_{\alpha \beta}}{2}
 \hbox{Tr}\left( \left( \gamma^3_{\theta^k} \otimes \gamma^{7\ph{g}*}_{\theta^k} + 
 \gamma^{3\ph{g}*}_{\theta^k} \otimes \gamma^7_{\theta^k} \right) \frac{i(\beta_1 + \beta_2)}{2 \pi^2}
\frac{\vartheta \Big[ \begin{array}{c} \alpha \\ \beta \end{array} \Big] \left(\frac{i 
\epsilon t}{2}\right)}{\vartheta \Big[ \begin{array}{c} 1/2 \\ 1/2 \end{array} \Big] 
\left(\frac{i 
\epsilon t}{2}\right) } \right)   \nonumber \\
& & 
\Bigg[ 
 \prod_{i=1}^{2} 
\frac{\vartheta \Big[ \begin{array}{c} 1/2 - \alpha \\ \beta + \theta_i \end{array} \Big]}
{\vartheta \Big[ \begin{array}{c} 0 \\ 1/2 + \theta_i \end{array} \Big] }
\left( - 2 \sin \pi \theta_3 \right)
\frac{\vartheta \Big[ \begin{array}{c} \alpha \\ \beta + \theta_3 \end{array} \Big]}
{\vartheta \Big[ \begin{array}{c} 1/2 \\ 1/2 + \theta_3 \end{array} \Big] } \Bigg].
\eea
The trace is over all D3-D7 and D7-D3 states, weighted by both the Chan-Paton matrices and the
effects of magnetic charges.

In this case
\bea
\hbox{Tr}\left( \left( \gamma_{\theta^3} \otimes \gamma^{-1}_{\theta^7} + \gamma_{\theta^7} \otimes \gamma^{-1}_{\theta^3} \right)
\frac{i(\beta_1 + \beta_2)}{2 \pi^2}
\frac{\vartheta \Big[ \begin{array}{c} \alpha \\ \beta \end{array} \Big] \left(\frac{i 
\epsilon t}{2}\right)}{\vartheta \Big[ \begin{array}{c} 1/2 \\ 1/2 \end{array} \Big] 
\left(\frac{i 
\epsilon t}{2}\right) } \right) & = & 
- \frac{B^2 t}{16 \pi^4} \ti 
\frac{\vartheta'' \Big[ \begin{array}{c} \alpha \\ \beta \end{array} \Big]}{\eta^3}
\Big( n_0^7 - \frac{n_1^7 + n_2^7}{2} \Big). \nonumber
\eea
Threshold corrections therefore take the form
\bea
\mc{A}_{37}^{(k)} & = & \frac{1}{3}\int \frac{dt}{2t} \left( \frac{B}{2 \pi^2} \right)^2 \frac{1}{8 \pi^2}
\Big( n_0^7 - \frac{n_1^7 + n_2^7}{2} \Big) \ti  \nonumber \\
& & 
\sum_{\alpha, \beta=0,1/2}  \frac{\eta_{\alpha \beta}}{2} 
\frac{\vartheta'' \Big[ \begin{array}{c} \alpha \\ \beta \end{array} \Big]}{\eta^3}
\Bigg[ 
 \prod_{i=1}^{2} 
\frac{\vartheta \Big[ \begin{array}{c} 1/2 - \alpha \\ \beta + \theta_i \end{array} \Big]}
{\vartheta \Big[ \begin{array}{c} 0 \\ 1/2 + \theta_i \end{array} \Big] }
\left( - 2 \sin \pi \theta_3 \right)
\frac{\vartheta \Big[ \begin{array}{c} \alpha \\ \beta + \theta_3 \end{array} \Big]}
{\vartheta \Big[ \begin{array}{c} 1/2 \\ 1/2 + \theta_3 \end{array} \Big] } \Bigg].
\eea
The expressions for the $\theta$ and $\theta^2$ twisted sectors are identical.
In the IR limit $t \to \infty$, we find
\be
\lim_{t \to \infty}
\sum_{\alpha, \beta=0,1/2}  \frac{\eta_{\alpha \beta}}{2} 
\frac{\vartheta'' \Big[ \begin{array}{c} \alpha \\ \beta \end{array} \Big]}{\eta^3}
\Bigg[ 
 \prod_{i=1}^{2} 
\frac{\vartheta \Big[ \begin{array}{c} 1/2 - \alpha \\ \beta + \theta_i \end{array} \Big]}
{\vartheta \Big[ \begin{array}{c} 0 \\ 1/2 + \theta_i \end{array} \Big] }
\left( - 2 \sin \pi \theta_3 \right)
\frac{\vartheta \Big[ \begin{array}{c} \alpha \\ \beta + \theta_3 \end{array} \Big]}
{\vartheta \Big[ \begin{array}{c} 1/2 \\ 1/2 + \theta_3 \end{array} \Big] } \Bigg] = -\pi^2(1 + \mc{O}(\sqrt{q}) + \ldots ).
\ee
The $t \to \infty$ contribution of the twisted sectors $(\theta + \theta^2)$ is therefore given by
\be
\label{d3d7betatwisted}
\left( -\int \frac{dt}{2t} \frac{1}{4} \left( \frac{B}{2 \pi^2} \right)^2 \right) \ti \frac{1}{3} \left(-n_0^7 + 
\frac{n_1^7 + n_2^7}{2} \right).
\ee
Combining (\ref{d3d7betatwisted}) and (\ref{d3d7betauntwisted}) we obtain
\be
\label{d3d7tot}
\left( -\int \frac{dt}{2t} \frac{1}{4} \left( \frac{B}{2 \pi^2} \right)^2 \right) \ti \left( 
\frac{n_1^7 + n_2^7}{2} \right).
\ee
This gives the correct contribution to the $\beta$ function from D3-D7 states, 
$b_{D3-D7} = \frac{n_1^7 + n_2^7}{2}$.

Combining all sectors by summing (\ref{33aiqu}) and (\ref{d3d7tot}), we obtain in the $t \to \infty$ limit the 
correct $\beta$ function coefficient for $SU(n_0)$,
\be
b_0 = -3n_0 + \frac{3(n_1 + n_2)}{2} + \frac{(n_1^7 + n_2^7)}{2}.
\ee 
The stringy thresholds exist in the UV $t \to 0$ limit. We can evaluate this limit either by transforming the
amplitudes (\ref{aaa}) and (\ref{gag}) to closed string channel and studying the $l \to 0$ limit or by direct numerical
evaluation. Either way, we find that the contribution of the twisted sector amplitudes vanishes in the $t \to 0$ limit:
$$
\lim_{t \to 0} \mc{A}^{33}_{\theta} - \mc{A}^{37}_{\theta} = \lim_{t \to 0} \mc{A}^{33}_{\theta^2} - \mc{A}^{37}_{\theta^2} = 0.
$$
This is consistent with the decoupling of $\mc{N}=1$ twisted sectors in the $t \to 0$ limit. This leaves the $\mc{N}=2$ amplitude
(\ref{d3d7betauntwisted}) associated to the D3-D7 untwisted sector. As the oscillator sum here reduces to a single number, this
amplitude is uncancelled and remains divergent in the $t \to 0$ limit,
\be
\left( -\int^{(\mu^2)^{-1}}_{(M^2)^{-1}} \frac{dt}{2t} \frac{1}{4} \left( \frac{B}{2 \pi^2} \right)^2 \right) \ti \frac{\left(n_0^7 + n_1^7 + n_2^7 %%@
\right)}{3}
= \left[ \frac{1}{8} \left( \frac{B}{2 \pi^2} \right)^2 \right] \ti \frac{\left(n_0^7 + n_1^7 + n_2^7 \right)}{3} \log (M^2/\mu^2).
\ee
This divergence is allowed as it originates from the untwisted sector: as for the 33 models, untwisted tadpoles
do not have to be cancelled locally but instead may propagate into the bulk, and be cancelled far from the local geometry.

However in this case the interpretation is not as clear as compared to the 33 examples. For this 37 example, the 
gauge couplings do all run above the string scale, but with a modified beta function coefficient
$b_0 = \frac{\left(n_0^7 + n_1^7 + n_2^7 \right)}{3}$. In contrast to the 33 cases, 
this is gauge-group universal and differs from the low-energy beta functions. As for the 33 examples, we expect this divergence to
be cut off at a scale $R M_s$ due to bulk tadpole cancellation.

It is also not easy to match this behaviour with the effective field theory. We could only find a match by making the following
assumptions
\begin{enumerate}
\item
We redefine the real part of the dilaton as
$$
\hbox{Re}(S) \to \hbox{Re}(S) - \frac{2}{9} \frac{1}{16 \pi^2} (n_0^7 + n_1^7 + n_2^7) \ln \mc{V}.
$$
\item 
The matter metric for D3-D7 matter is $Z^{37} = \frac{1}{\mc{V}}$, rather than $Z^{33} = \frac{1}{\mc{V}^{2/3}}$.
\end{enumerate}
However there are two difficulties with this interpretation. First, it reinvolves a redefinition of the dilaton
by an amount depending on the number of D7 brane stacks. However, the dilaton also provides the gauge kinetic function
for D3 brane stacks far from the singularity. As such stacks need have no light D3-D7 strings there seems no reason the gauge
coupling on such stacks should be sensitive to the number of D7 branes present. Secondly, writing $Z^{37} = \frac{1}{\mc{V}}$ implies
the physical Yukawas for (37)(73)(33) couplings diverge as $\mc{V}^{1/3}$. This is inconsistent with the locality of the model, rendering it
impossible to take the $\mc{V} \to \infty$ limit that is essential to a local model. 

For these reasons it does not seem that this interpretation is correct. We believe the reason we have found difficulty matching
with the effective field theory is that D3-D7 models are not truly local and weakly coupled, due to the large backreaction of
D7 branes on spacetime: D7 branes source a divergence for the dilaton, which in turn sets the gauge coupling on
probe D3 branes. 
In this case the $\mc{V} \to \infty$ limit is not a well-defined limit due to the existence of a D7 brane extending out 
into the bulk, and the basic assumptions that we used in section \ref{secFT} are not valid. 

\section{Conclusions}

This paper has studied threshold corrections for local string models embedded in a large compact space.
We showed that for such models the Kaplunovsky-Louis anomaly formula for 
effective field theory implies threshold corrections
should modify the unification scale from $M_s$ to $R M_s$ due to the Konishi and
super-Weyl anomalies.

We also analysed this issue from a directly stringy perspective and found full agreement with the KL formula
for a large class of models,
namely those arising from fractional D3 branes at orbifold singularities. From a string perspective the open 
string loop diagram entering the threshold corrections can be reinterpreted as a closed string tree diagram. 
This diverges in the local model due to the emission of a RR tadpole into 
the bulk. In homology this corresponds to charge along a cycle whose global homology status 
cannot be determined locally.
In a consistent compact model this tadpole must be cancelled in the bulk at a 
distance $R l_s$ from the local channel. In open string 
channel this effectively regularises the divergence of the local model through new charged states of 
mass $R M_s$. At energies below $R M_s$ the gauge couplings start running as these
states decouple.

For D3/D7 models we found a universal running above the string scale, which however differed from 
the low energy beta functions.
However in this case we were unable to match onto the KL field theory formula. While we do not fully understand 
the origin of this discrepancy,
it may be due to our omitting the large backreaction from D7 branes - truly 
local D3/D7 models may not exist.

An obvious future direction is to extend this study of threshold corrections to more general models. This 
includes both orientifolded singularities and models in the geometric regime away from the orbifold limit -
this includes for example local models of intersecting D7 branes.
The field theory formula seems surprisingly general in its implications, and from a string theory point of view it would
be very interesting to analyse carefully the issue of threshold corrections for general local models.

Another direction to analyse is the effect that resolving the singularity has on the gauge couplings. Our treatment
here has been carried out in the orbifold limit where the string computation is valid. Resolving the singularity may 
lead to a tree-level splitting of the gauge couplings. This could further raise the unification scale above the string scale,
or alternatively return the unification scale to the string scale. This issue deserves further study.

Finally, although this paper has focussed on the technical calculational details, its motivation was phenomenological:
what is the significance of apparent gauge coupling unification at $M_{GUT} \sim 10^{16} \hbox{GeV}$? 
The results here imply that models with $M_s < M_{GUT}$ are not \emph{a priori} incompatible with gauge coupling unification; indeed,
for local models the string scales and unification scales may differ substantially.
This has clear implications for proton decay: if the string scale for local GUT constructions is $\sim 10^{15} \hbox{GeV}$, proton
decay is substantially enhanced over more conventional estimates. 

For models with intermediate string scales $M_s \sim 10^{11} \hbox{GeV}$, which are
most attractive in terms of generating low-scale supersymmetry and solving the hierarchy problem, the unification scale becomes
$10^{13} \hbox{GeV} \div 10^{14} \hbox{GeV}$. This significantly ameliorates but does not eliminate the
tension between the intermediate and GUT scales. For such models to be consistent with gauge coupling unification, a certain amount
of non-universality in the tree level gauge couplings may be necessary.

\acknowledgments{This work was initiated at the IPMU in Tokyo, who I thank for their hospitality.
I also thank Graham Ross for not allowing me to forget gauge coupling unification. 
I have learned from discussions with Shanta de Alwis, Florian Gmeiner,  
Mark Goodsell, Xenia de la Ossa, Fernando Quevedo, Eran Palti, Bert Schellekens, 
James Sparks and Taizan Watari, and
I thank Fernando Quevedo and Luis Ibanez for comments on the paper. I 
am grateful to
the Royal Society who kindly support me with a University Research Fellowship. } 

\appendix

\section{$\vartheta$-function identities}

We here collate definitions and identities of the various Jacobi-$\vartheta$ functions.
We write $q = e^{-\pi t}$ throughout these formulae. The eta function is defined by
\be
\eta(t) = q^{1/24} \prod_{n=1}^{\infty}(1 - q^n).
\ee
The Jacobi $\vartheta$-functon with general characterstic is defined as
\be
\vartheta \Big[ \begin{array}{c} \alpha \\ \beta \end{array} \Big](z | t) = 
\sum_{n \in \, \mbb{Z}} e^{- (n + \alpha)^2 \pi t/2} e^{2 \pi i (z + \beta) (n + \alpha)}.
\ee 
Here $z = 0$ unless specified. 
The $\vartheta$ functions are manifestly invariant under $\alpha \to \alpha + \mbb{Z}$.
A useful expansion valid for $\alpha \in (-\half, \half]$ is
\be
\frac{\vartheta \Big[ \begin{array}{c} \alpha \\ \beta \end{array} \Big]}{\eta}(t)
= e^{2 \pi i \alpha \beta} q^{\frac{\alpha^2}{2} - \frac{1}{24}} \prod_{n=1}^{\infty}
\left( 1 + e^{2 \pi i \beta} q^{n- \half + \alpha} \right) \left(1 + e^{-2 \pi i \beta} q^{n-\half -\alpha}  \right). 
\ee
For the four special $\vartheta$-functions, we have
\bea
\vartheta_1(z | t) \equiv \vartheta \Big[ \begin{array}{c} \half \\ \half \end{array} \Big](z | t)
& = & 2 q^{1/8} \sin \pi z \prod_{n=1}^{\infty} (1 - q^n) (1-e^{2 \pi iz} q^n) (1 - e^{-2 \pi i z} q^n). \\
\vartheta_2(z | t) \equiv \vartheta \Big[ \begin{array}{c} \half \\ 0 \end{array} \Big](z | t)
& = & 2 q^{1/8} \cos \pi z \prod_{n=1}^{\infty} (1 - q^n) (1 + e^{2 \pi iz} q^n) (1 + e^{-2 \pi i z} q^n). \\
\vartheta_3(z | t) \equiv \vartheta \Big[ \begin{array}{c} 0 \\ 0 \end{array} \Big](z | t)
& = &  \prod_{n=1}^{\infty} (1 - q^n) (1 + e^{2 \pi iz} q^{n-\half}) (1 + e^{-2 \pi i z} q^{n-\half}). \\
\vartheta_4(z | t) \equiv \vartheta \Big[ \begin{array}{c} 0 \\ \half \end{array} \Big](z | t)
& = &  \prod_{n=1}^{\infty} (1 - q^n) (1 - e^{2 \pi iz} q^{n-\half}) (1 - e^{-2 \pi i z} q^{n-\half}). 
\eea
These appear in the partition functions of a magentised sector. We shall normally leave the 
$t$ argument implicit when using these. Derivatives w.r.t $z$ give
\bea
\vartheta_1(z) & = & 2 \pi \eta^3 z + \mc{O}(z^3), \\
\vartheta_i(z) & = & \vartheta_i(0) + \frac{z^2}{2} \vartheta_i^{''}(0) + \mc{O}(z^4), \qquad i = 2,3,4.
\eea
$\mc{N}=1$ expressions for threshold corrections can be simplified using the identities
\bea
\label{ts1}
\sum_{\alpha, \beta} \eta_{\alpha, \beta} \frac{\vartheta'' \Big[ \begin{array}{c} \alpha \\ \beta \end{array} \Big]}{\eta^3}
\prod_{i=1}^3 
\frac{\vartheta \Big[ \begin{array}{c} \alpha \\ \beta + \theta_i \end{array} \Big]}
{\vartheta \Big[ \begin{array}{c} 1/2 \\ 1/2 + \theta_i \end{array} \Big] } & = &
\label{ts2}
-2 \pi \sum_{i=1}^3 \frac{\vartheta' \Big[ \begin{array}{c} 1/2 \\ 1/2 - \theta_i \end{array} \Big]}
{\vartheta \Big[ \begin{array}{c} 1/2 \\ 1/2 - \theta_i \end{array} \Big] }, \\
\lim_{t \to \infty} \frac{\vartheta' \Big[ \begin{array}{c} 1/2 \\ 1/2 - \theta_i \end{array} \Big]}
{\vartheta \Big[ \begin{array}{c} 1/2 \\ 1/2 - \theta_i \end{array} \Big] }
& = & \frac{ \pi \cos (\theta_i)}{\sin (\theta_i)}.
\eea
$\mc{N}=2$ threshold corrections are much simplified using the result
\be
\label{ts3}
\sum \eta_{\alpha \beta} (-1)^{2 \alpha}  
\frac{\vartheta'' \Big[ \begin{array}{c} \alpha \\ \beta \end{array} \Big]}{\eta^3 }  
\frac{\vartheta \Big[ \begin{array}{c} \alpha \\ \beta \end{array} \Big]}{\eta^3 }  
\frac{\vartheta \Big[ \begin{array}{c} \alpha \\ \beta + \theta_1  \end{array} \Big]}
{\vartheta \Big[ \begin{array}{c} 1/2 \\ 1/2 + \theta_1 \end{array} \Big] }
\frac{\vartheta \Big[ \begin{array}{c} \alpha \\ \beta + \theta_2  \end{array} \Big]}
{\vartheta \Big[ \begin{array}{c} 1/2 \\ 1/2 + \theta_2 \end{array} \Big] } = - 4\pi^2
\ee
for $\theta_1 + \theta_2 = 1 \hbox{ mod } 2$.

When transforming to closed string channel the following identities are useful
\bea
\label{EtaTransform}
\eta(t) & = & \sqrt{l} \eta(l), \\
\label{ThetaTransform}
\vartheta \Big[ \begin{array}{c} \alpha \\ \beta \end{array} \Big](t) & = & \sqrt{l} e^{2 \pi i \alpha \beta}  
\vartheta \Big[ \begin{array}{c} -\beta \\ \alpha \end{array} \Big](l).
\eea

\end{document}